\documentclass[12pt]{iopart}

\expandafter\let\csname equation*\endcsname\relax
\expandafter\let\csname endequation*\endcsname\relax

\pdfoutput=1

\usepackage[T1]{fontenc}
\usepackage{graphicx}
\usepackage{caption}
\usepackage{subcaption}
\usepackage{diagbox}
\usepackage{color}
\usepackage[backref=none]{hyperref}
\hypersetup{
  colorlinks,
  citecolor=red,
  filecolor=black,
  linkcolor=blue,
  urlcolor=magenta
}
\usepackage{amsmath}
\usepackage{amssymb}
\usepackage{mathtools}
\usepackage{lineno}
\usepackage{cite}

\numberwithin{equation}{section} 
\numberwithin{figure}{section} 

\allowdisplaybreaks[1]

\newcounter{tempcounter}

\newcommand{\sfrac}[2]{{}^{#1}\!\!/{}\!_{#2}}
\newcommand{\deltad}{\,\delta_\textsc{d}\!}
\newcommand{\thetah}{\,\Theta_\textsc{h}\!}
\newcommand{\xiN}{\underline{\xi}_{N}}
\newcommand{\PN}[2][\textsc{ne}]{P_{#1}^{(#2)}}
\newcommand{\avg}[1]{\langle #1 \rangle_{\textsc{ne}}}
\newcommand{\Ka}[1][\alpha]{K_{#1}}
\newcommand{\Ja}[1][]{J^{\alpha_{#1}}}
\newcommand{\hL}{\widehat{\mathcal{L}}}

\newcommand{\Jh}{J_{\textsc{h}}}

\begin{document}

\title
[Heat conduction of stochastic energy exchange processes]
{Heat conduction and the nonequilibrium stationary states of
  stochastic energy exchange processes } 

\author{Thomas Gilbert} 

\address{
  Center for Nonlinear Phenomena and Complex Systems,
  Universit\'e Libre  de Bruxelles, C.~P.~231, Campus Plaine, B-1050
  Brussels, Belgium
}

\date{Version of \today}

\begin{abstract}
  I revisit the exactly solvable Kipnis--Marchioro--Presutti  model of
  heat conduction [J. Stat. Phys. \textbf{27} 65 (1982)] and describe, for
  one-dimensional systems of arbitrary sizes whose ends are in contact
  with thermal baths at different temperatures, a systematic
  characterisation of their non-equilibrium stationary states. These
  arguments avoid resorting to the analysis of a dual process and
  yield a straightforward derivation of Fourier's law, as well as
  higher-order static correlations, such as the covariant matrix. The
  transposition of these results to families of gradient models
  generalising the KMP model is established and specific cases are
  examined.
\end{abstract}

\section{Introduction \label{sec:intro}}

The Kipnis--Marchioro--Presutti (KMP) model of heat conduction
\cite{Kipnis:1982Heat} consists of a one dimensional chain of harmonic
oscillators which exchange energy among nearest neighbours through
stochastic interactions. Considering a finite-size chain in contact
with thermal reservoirs at different temperatures, the existence and
unicity of the stationary measure was established, proving, in the
infinite system-size limit, the convergence of the distribution of
energies to a product measure of exponential Gibbs distributions whose
temperatures interpolate linearly between the two baths' temperatures. 
Moreover, the associated heat flux is proportional to the temperature
gradient, with uniform coefficient specifying the heat conductivity. 

In many respects, the KMP model is a prototypical example of a system
of interacting particles on a lattice which exhibits normal 
heat conduction and is amenable to an exact solution. 
In particular, it is one among a few models whose large deviation
functional is known, providing a characterisation of macroscopic
energy fluctuations about the non-equilibrium stationary 
state \cite{Bertini:2005Large}.

A potent tool which has proven very useful in establishing the
properties of the KMP model is that of duality
\cite{Liggett:2012interacting}, which allows to reduce the  
analysis of the model under consideration to that of a lattice gas of
particles randomly hopping and mixing among neighbouring sites until
they are absorbed at the boundaries; see also
reference~\cite{Bertini:2005Large}. The notion that the $r$-point
correlation functions in the nonequilibrium steady state can be
obtained from absorption probabilities of $r$ dual particles has in
the recent years led to a number of other fruitful applications of
duality in the context of  interacting particle systems
\cite{Giardina:2007Duality, Giardina:2009Duality,
  Carinci:2013Duality, Carinci:2015Dualities, Carinci:2016Asymmetric,
  vanGinkel:2016Duality}.

The objective pursued in this paper is to show that the
nonequilibrium stationary state of the KMP process can in fact be
fully characterised rather easily and without resorting to
duality. The technique is based on a straightforward expansion of the 
stationary state in terms of orthogonal polynomials whose coefficients
specify, for a given degree $r$, the $r$-point correlation
functions. In so doing, we recover the results already obtained by
Bertini \emph{et al.} \cite{Bertini:2007Stochastic} for the two-point
correlation function. Our technique is, however, more general and
yields, for systems of any size, a systematic derivation
of the coefficients of arbitrary degree, $r$. They are indeed
determined through a closed set of linear equations which involve
coefficients of degree $r'\leq r$.  This is of course consistent with
the applicability of duality and is indeed key to its usefulness. Yet
the simplicity and straightforwardness of the approach described here
seems not to have been duly appreciated.

A remarkable property of the KMP model is that the deterministic part
of the current associated with two neighbouring cells with fixed
energies, i.e.~the first moment of the KMP stochastic kernel
associated with its generator, is proportional to the difference of
their energies and is therefore similar in form to its thermodynamic
counterpart, given by Fourier's law. This property, which is known as
the gradient property \cite{Spohn:1991book}, is central to the model's
simplicity and the fact that its transport coefficient is given in
terms of the current's average value with respect to local thermal
equilibria. Further models 
of heat transport similar to the KMP process in that they share the
gradient property have been considered in recent years; see
references~\cite{Giardina:2005Fourier, Giardina:2009Duality,
  Cirillo:2014Duality, Redig:2015Multilinearity, 
  Nandori:2016Local, Franceschini:2017stochastic}. Of particular
interest for our sake are  so-called Brownian energy processes (BEP)
\cite{Giardina:2009Duality}, which have been extensively studied in
the framework of duality \cite{Grosskinsky:2011Condensation}; see also
references~\cite{Giardina:2010Correlation} and \cite{Peletier:2014Large}.  

Here, we consider a family of models derived from the instantaneous
thermalisation regime of BEP \cite{Giardina:2009Duality,
  Franceschini:2014energy}, whose local equilibrium distributions are
specified in terms of Gamma distributions with arbitrary shape
parameters, $\alpha>0$ (the local temperatures specify the scale
parameters). Gamma distributions with half integer shape parameters, 
$\alpha = n/2$, are typically encountered when considering the energy
distributions of mechanical systems of $n$ particles  \cite[Chapter
1]{vanKampen:2007Book}. Our analysis is, however, not constrained by
such restrictions; we consider positive real valued shape parameters
and, in general, let them take different values in different cells. We
show the nonequilibrium stationary states of such systems can in fact
be characterised in the same way as that of the KMP process,
irrespective of the configuration of shape parameters, wether uniform,
periodic or disordered. The consequences of configurational disorder
on heat conduction are usually investigated in the framework of
harmonic chains \cite{Lepri:2003Thermal}. Our models provide a
different class and are amenable to analytic results.

The problem of characterising the nonequilibrium stationary states of
such systems may however be quite complex since, in general, the
temperature profile does not have a simple linear form. On the one
hand, periodic shape parameter configurations are quite easy to treat,
at least as far as the computation of the heat conductivity goes: it
is proportional to the harmonic mean of the shape
parameters. Disordered configurations of shape parameters may, on the
other hand, lead to pathological cases, in particular, when the shape
parameters can have arbitrarily small values. We thus focus 
more specifically on the study of systems with uniform and alternating
shape parameters, which all have linear temperature profiles. An
example of such a system was recently considered in
reference~\cite{Nandori:2016Local}, corresponding to shape parameters 
alternating between $1$ and $1/2$. There, its dual was shown to be a
symmetric simple exclusion process with alternating jump rates
\cite{Richards:1977Theory, Wick:1989Hydrodynamic}; see also
references \cite{Faggionato:2003Hydrodynamic,  Quastel:2006Bulk} for
disordered cases. The linear temperature profile of the energy
exchange processes allows us to obtain their two-point correlation
functions. In the case of uniform shape parameters, we briefly discuss
two limiting cases, corresponding respectively to small and large
shape parameter values. In the former case, the limiting regime is
such that indivisible energy packets perform random walks in the
vicinity of empty (zero energy) cells and merge whenever two energy
packets cross each other, which is similar to diffusion-aggregation
models \cite{Spouge:1988Exact, Thomson:1989Exact}. The latter limits
to a process where the total energy of the interacting cells is halved
evenly among them. 

The paper is organised as follows. The general features and properties
of the KMP model are recalled in \sref{sec:KMP}. In \sref{sec:ness},
we describe the construction of its stationary state by a polynomial
expansion in the energy variables, focusing in sections
\ref{sec:KMPdeg1} and \ref{sec:KMPdeg2} respectively on first and
second degree contributions for which general solutions are easily
inferred for any system size. In \sref{sec:KMPN0}, we consider the
special case of a single-cell system and compare the results of our
approach to the exact solution found in
reference~\cite{Bertini:2007Stochastic}. The case of a two-cell system
is then considered in \sref{sec:KMPN1} which allows to show that
explicit solutions can easily be found for arbitrary degrees of the
polynomial expansion of the stationary state. The extension of the KMP
model to families of such models specified by a configuration of
positive real 
shape parameters is discussed in \sref{sec:KMPalpha}. In
\sref{sec:KMPalphacurrent} general results are obtained for the first
degree terms of the polynomial expansion of the stationary states of
such models and different models are discussed. The second degree
coefficients are  derived in \sref{sec:KMPalphauniform} and
\sref{sec:KMPalpha01} for uniform and alternating shape parameters
respectively. Conclusions are drawn in \sref{sec:con}. The appendices
provide a number of technical details pertaining to polynomial
expansions of the stationary states. 

\section{The Kipnis--Marchioro--Presutti model \label{sec:KMP}} 

In the original KMP model \cite{Kipnis:1982Heat} the state of a system
of $N+1$ cells on a one-dimensional lattice is specified by a collection
$\xiN \equiv\{\xi_{\sfrac{-N}{2}}, \dots, \xi_{\sfrac{N}{2}}\}$ of
positive real variables, $\xi_{i} \in \mathbb{R}_{+}$, interpreted as
energies, which are let to interact pairwise 
through the stochastic kernel
\begin{equation}
  \label{eq:KMPkernel}
  K(\xi_{a}, \xi_{b}\rightarrow \xi'_{a}, \xi'_{b}) =  
  \frac{\nu}{\xi_{a} + \xi_{b}} 
  \deltad( \xi_{a} + \xi_{b} - \xi'_{a} - \xi'_{b}) 
  \thetah(\xi'_{a}) \thetah(\xi'_{b}) \delta_{|a-b|,1}\,,
\end{equation}
i.e.~such that the combined energy $\xi_{a} + \xi_{b}$ of the
interacting nearest neighbouring cells $a$ and $b$ is uniformly
redistributed among themselves. 

This amounts to picking a pair $\{n,n+1\}$ with uniform rate $\nu$
and drawing a uniformly distributed random number $p \in (0,1)$ such
that 
\begin{equation}
  \label{eq:xipxi}
  \begin{split}
   \xi'_{n} &= p\, (\xi_{n} + \xi_{n+1})\,,\\ 
   \xi'_{n+1} &= (1 - p) (\xi_{n} + \xi_{n+1})\,.
  \end{split}
\end{equation}
Furthermore, the kernel \eqref{eq:KMPkernel} satisfies the detailed
balance condition,
\begin{equation}
  \label{eq:detailedbalance}
  \PN[\textsc{eq}]{N}(\dots,\xi_{a}, \xi_{b}, \dots)
  K(\xi_{a}, \xi_{b}\rightarrow \xi'_{a}, \xi'_{b})
  = 
  \PN[\textsc{eq}]{N}(\dots, \xi'_{a}, \xi'_{b}, \dots)
  K(\xi'_{a}, \xi'_{b} \rightarrow \xi_{a}, \xi_{b})
  \,,
\end{equation}
with microcanonical equilibrium distribution specified by the
condition $\xi_{\sfrac{-N}{2}} +  \dots + \xi_{\sfrac{N}{2}} =
\beta^{-1}(N+1)$, 
which, as $N\to\infty$, tends to the product of canonical
distributions with inverse temperature $\beta$, 
$\PN[\textsc{eq}]{N}(\xiN) = \prod_{a} p_{\beta}(\xi_{a})$,
\begin{equation}
  \label{eq:canonicaldist}
  p_{\beta}( \xi) = \beta \, \rme^{-\beta \, \xi} 
  \,.
\end{equation}

Beyond the uniform rate, identified with the zeroth moment of the 
kernel \eqref{eq:KMPkernel},
\begin{equation} 
  \label{eq:KMPfrequency}
  f(\xi_1, \xi_2)  = \int \rmd \xi'_{1} \, \rmd\xi'_{2}\, 
  K(\xi_{1}, \xi_{2}\rightarrow \xi'_{1}, \xi'_{2}) = \nu \,,
\end{equation}
the higher moments are
\begin{equation}
  \label{eq:KMPkernelmoments}
  \int \rmd \xi'_{1} \, \rmd\xi'_{2}\, (\xi_{1} - \xi'_{1})^n 
  K(\xi_{1}, \xi_{2}\rightarrow \xi'_{1}, \xi'_{2}) = 
  \frac{\nu}{n+1} \frac{\xi_1^{n+1} + (-)^n \xi_2^{n+1}} {\xi_1 + \xi_2}
  \,.
\end{equation}
In particular, the first moment defines the current,
\begin{equation}
  \label{eq:KMPcurrent}
  j(\xi_1, \xi_2) = \frac{\nu}{2} (\xi_1 - \xi_2)\,,
\end{equation}
which is the gradient of the local energies, a simple instance of the
gradient property; see reference~\cite{Spohn:1991book}. As a 
consequence, the coefficient of heat  conductivity is simply
\begin{equation}
  \label{eq:KMPconductivity}
  \kappa = \frac{\nu}{2}
\end{equation}
In other words, the infinite system-size
limit of the heat current is linear in the local temperature gradient
and, up to a sign, the coefficient of proportionality is $\kappa$,
which is uniform throughout. 

This property actually holds also for systems of finite sizes in a
nonequilibrium stationary state associated with a temperature
gradient. In the following section, we consider systems of arbitrary
sizes whose ends are in contact with thermal baths at different
temperatures and obtain a systematic characterisation of their
stationary state in terms of the energy moments. In particular, we
obtain equation \eqref{eq:KMPconductivity} from the degree-$1$
contributions.

\section{Non-equilibrium stationary state \label{sec:ness}}

To drive the system away from equilibrium, one lets the two ends of
the system be coupled to thermal baths with respective inverse 
temperatures $\beta_{\pm}$ and distributions $p_{\beta_{\pm}}$,
as specified by equation \eqref{eq:canonicaldist}. Other
choices may, however, be more convenient; see
equation \eqref{eq:thermalbathalt} below.  

The time evolution of the distribution $\PN[t]{N}(\xiN)$ thus satisfies the
following master equation,
\begin{equation}
  \label{eq:KMPmeq}
  \partial_t \PN[t]{N}(\xiN) 
  = 
  \sum_{a = \sfrac{-N}{2}}^{\sfrac{N}{2} - 1}
  \hL_{a,a+1} \PN[t]{N}(\xiN)  
  + 
  \hL_{\sfrac{-N}{2}} \PN[t]{N}(\xiN)  
  + 
  \hL_{\sfrac{N}{2}} \PN[t]{N}(\xiN) \,, 
\end{equation}
where the local exchange operators are self-adjoint operators,
defined by 
\begin{multline}
  \label{eq:KMPLP}
  \hL_{a,a+1} \PN[t]{N}(\{\dots, \xi_{a}, \xi_{a+1},\dots\})
  = 
  \cr
  \int \rmd\xi'_a \, \rmd \xi'_{a+1} \,
  K(\xi'_{a}, \xi'_{a+1}\rightarrow \xi_{a}, \xi_{a+1})
  \PN[t]{N}( \{\dots, \xi'_{a}, \xi'_{a+1},\dots\} )
  - \nu \PN[t]{N}(\xiN)\,,
\end{multline} 
and the thermal boundary conditions are specified, on the left-hand
boundary, by the operator
\begin{multline}
  \label{eq:KMPLPleft}
  \hL_{\sfrac{-N}{2}} \PN[t]{N}(\{\xi_{\sfrac{-N}{2}},\dots\} )  
  =
  \cr
  \int \rmd\xi_{-} \, \rmd\xi'_{-} \, \rmd\xi'_{\sfrac{-N}{2}} \,
  K(\xi'_{-}, \xi'_{\sfrac{-N}{2}}\rightarrow \xi_{-}, \xi_{\sfrac{-N}{2}})
  p_{\beta_{-}}( \xi'_{-}) 
  \PN[t]{N}( \{\xi'_{\sfrac{-N}{2}},\dots\} )
  - \nu \PN[t]{N}(\xiN)\,,
\end{multline} 
and similarly for right-hand boundary,
\begin{multline}
  \label{eq:KMPLPright}
  \hL_{\sfrac{N}{2}} \PN[t]{N}(\{\dots,\xi_{\sfrac{N}{2}}\} )  
  =
  \cr
  \int \rmd\xi_{+} \, \rmd \xi'_{+} \, \rmd \xi'_{\sfrac{N}{2}}\,
  K(\xi'_{\sfrac{N}{2}}, \xi'_{+}\rightarrow \xi_{\sfrac{N}{2}}, \xi_{+})
  p_{\beta_{+}}( \xi'_{+}) 
  \PN[t]{N}( \{\dots, \xi'_{\sfrac{N}{2}}\})
  - \nu \PN[t]{N}(\xiN)\,.
\end{multline}
In these expressions the dots in the arguments of the distributions on
the right-hand 
side stand for the components of $\xiN$ identical to those appearing
on the left-hand side.

The non-equilibrium stationary state $\PN{N}$ of this process can be
conveniently expanded in terms of Laguerre polynomials, which form an
orthonormal set with respect to the weight function
\eqref{eq:canonicaldist} and are the natural choice for this model:
\begin{equation}
  \label{eq:KMPness}
  \PN{N} (\xiN) = 
  \!\!
  \prod_{a = \sfrac{-N}{2}}^{\sfrac{N}{2}} 
  \!\!
  p_{\beta_a}(\xi_{a})
  \!\!\!\!\sum_{n_{\sfrac{-N}{2}}, \dots, n_{\sfrac{N}{2}} = 0}^{\infty}  
  \gamma_{n_{\sfrac{-N}{2}} \dots n_{\sfrac{N}{2}}}
  L_{n_{\sfrac{-N}{2}}}(\beta_{\sfrac{-N}{2}} \xi_{\sfrac{-N}{2}}) \dots
  L_{n_{\sfrac{N}{2}}}(\beta_{\sfrac{N}{2}} \xi_{\sfrac{N}{2}})
  \,.
\end{equation}
We are thus faced with the problem of determining the parameters
$\beta_n$ and coefficients $\gamma_{n_{\sfrac{-N}{2}} \dots
  n_{\sfrac{N}{2}}}$ in equation \eqref{eq:KMPness}.  Given a set of
indices $\{n_{\sfrac{-N}{2}}, \dots, n_{\sfrac{N}{2}}\}$, we refer to
the sum of the indices $r = \sum_{i} n_i$  as the degree of the
coefficients. We show below that these coefficients can be determined
by sets of equations which  are closed, degree by degree in the sense
that terms of degree $r$ are determined through a set of linear equations
involving coefficients of degree less than or equal to
$r$. Coefficients of degree $r$ can therefore be determined exactly,
without having to invoke a closure approximation. With hindsight, one
realises this property is key to the applicability of duality to the
KMP process.

Prior to this, however, we note that, by normalisation of the stationary
distribution, we must have:
\begin{equation}
  \label{eq:gamma0}
  \gamma_{0,\dots,0} = 1\,.  
\end{equation}
Moreover, requiring that the parameters $\beta_n$ are the
inverse temperatures\footnote{Here and in the sequel, we denote the
  integration with respect to stationary state \eqref{eq:KMPness} by
  $\avg{\cdot}$. }, i.e.~$\avg{\xi_n} = \beta_n^{-1}$, imposes, for
all $n$, 
\begin{equation}
  \label{eq:gamma1}
  \gamma_{0,\dots,0,\underbracket[0.5pt][2pt]{\scriptstyle 1}_{n},0,\dots,0} = 0\,.  
\end{equation}
Of course, this does not tell us what are the values of the
parameters $\beta_n$, which we turn to below.

For ease of notation, we will denote the above elements by
$\gamma_{n:1}$, meaning that the $n$\textsuperscript{th} index is $1$
and all the others are zero. This notation extends to other
combinations of indices in a self-explanatory way so that,
e.g.~degree-$2$ coefficients correspond to all combinations of
$\gamma_{n:2}$ and $\gamma_{n:1,m:1}$, $n < m$.

\subsection{Degree-$1$ contributions: heat
  current \label{sec:KMPdeg1}} 

It is an immediate consequence of equations
\eqref{eq:gamma0}-\eqref{eq:gamma1} that the stationary expectation
value of the current \eqref{eq:KMPcurrent} is 
\begin{equation}
  \label{eq:KMPFourier}
  \avg{ j(\xi_n, \xi_{n+1}) } = \frac{\nu}{2} (\beta_{n}^{-1} -
  \beta_{n+1}^{-1})\,.
\end{equation}
Provided the parameters $\beta_n$ correspond to inverse local
temperatures (and their pairwise differences to local temperature
gradients), the above equation establishes Fourier's law of heat
conduction with uniform heat conductivity
\eqref{eq:KMPconductivity}. Indeed, let $\Jh(N+2) = \sum_{n} \avg{
  j(\xi_n, \xi_{n+1}) }$ denote the total of the average
currents through the $N+2$ energy pairs (including thermal
baths). Equation \eqref{eq:KMPFourier} implies $\Jh(N+2) = 
-\tfrac{1}{2}\nu(\beta_{+}^{-1}  - \beta_{-}^{-1})$. Furthermore,
since the current is constant throughout the system, we must therefore
have, for every pair of cells,
\begin{equation}
  \label{eq:KMPFourierlocal}
    \avg{ j(\xi_n, \xi_{n+1}) } = - \frac{\nu}{2} \frac{\beta_{+}^{-1} -
  \beta_{-}^{-1}}{N+2}\,.
\end{equation}
It thus remains to show the local temperature gradients are uniform
and given by the ratio of the temperature difference of the thermal
baths, $\beta_{+}^{-1} -  \beta_{-}^{-1}$, by the total number of
energy pairs, $N+2$. 

To determine the relationship between the parameters $\beta_n$ and the
baths inverse temperatures $\beta_{\pm}$, we consider the stationarity
of the first energy moments,
\begin{equation}
  \label{eq:1stenergymoment}
  \partial_t \avg{\xi_n} = 0\,,
\end{equation}
which is a short-hand notation for the multiplication of the
right-hand side of equation \eqref{eq:KMPmeq} by $\xi_n$ and
integrated over all variables $\xi_{\sfrac{-N}{2}}, \dots, \xi_{\sfrac{-N}{2}}$.
Its contributions arise from only two terms in equation
\eqref{eq:KMPmeq}, namely those representing the interaction of cell
$n$ with cell $n-1$ ($a = n-1$) and cell $n+1$ ($a=n$). The
contributions to each of these two terms come about from terms in
equation \eqref{eq:KMPness} with coefficients $\gamma_{n;p,n+1:q}$ and 
$\gamma_{n-1:q, n;p}$, i.e.~such that every index is zero except for the
$n$\textsuperscript{th} and $n\pm1$\textsuperscript{th} indices. For
each such term, we have 
\begin{multline}
  \label{eq:KMPbetanfromdt}  
  \int\rmd\xi_{n} \, \rmd\xi_{n\pm1}\,
  \beta_{n} \beta_{n\pm1} 
  \frac{\xi_{n}}{\xi_{n} + \xi_{n\pm1}} 
  \int_{-\xi_{n}}^{\xi_{n\pm1}}\rmd \eta \,
  \rme^{-\beta_{n}(\xi_{n} + \eta) - \beta_{n\pm1} (\xi_{n\pm1} - \eta)}    
  \cr
  \times
  L_{p}(\beta_{n}(\xi_{n} + \eta))
  L_{q}(\beta_{n\pm1} (\xi_{n\pm1} - \eta))    
  \cr
  = \tfrac{1}{2}(\beta_{n}^{-1} + \beta_{n\pm1}^{-1})
  \delta_{p,0} \, \delta_{q,0}
  - \tfrac{1}{2}\beta_{n}^{-1} \delta_{p,1} \, \delta_{q,0}
  - \tfrac{1}{2}\beta_{n\pm1}^{-1} \delta_{p,0} \, \delta_{q,1}
  \,;
\end{multline}
see \ref{app:1st} for details.
Therefore only the terms of degrees $0$ and $1$ in the stationary
state \eqref{eq:KMPness} contribute to the first energy moments.
Moreover, from equations \eqref{eq:gamma0} and \eqref{eq:gamma1}, we
see that only the first of the three terms on the right-hand side of 
equation \eqref{eq:KMPbetanfromdt} is actually relevant. Adding the
contributions from the two pairs of cells and subtracting
$2\beta_{n}^{-1}$, which arises from the loss term in
\eqref{eq:KMPLP}, we obtain  
\begin{equation}
  \label{eq:KMPbetan}
  \beta_{n}^{-1} = \tfrac{1}{2} ( \beta_{n+1}^{-1} + \beta_{n-1}^{-1} )\,.
\end{equation}
In terms of the baths' inverse temperatures, this implies
\begin{equation}
  \label{eq:KMPbetanN}
  \beta_{n}^{-1} = \tfrac{1}{2} ( \beta_{+}^{-1} + \beta_{-}^{-1} ) 
  + \frac{n}{N+2} ( \beta_{+}^{-1} - \beta_{-}^{-1} ) \,,
\end{equation}
i.e.~the stationary state is characterised by a linear profile of
temperatures interpolating between the temperatures of the two
baths.

The temperature gradient is therefore uniform across the system, so
that equation \eqref{eq:KMPFourierlocal} expresses the direct
proportionality of the local current and local temperature gradient,
with constant $\nu/2$ identified as the thermal conductivity.
In particular, for $N\to\infty$, we recover the linear temperature
profile obtained in reference~\cite{Kipnis:1982Heat} by other
methods. Correspondingly, we expect the stationary state to converge
to a product measure so that all coefficients $\gamma$ in 
\eqref{eq:KMPness} must vanish in this limit, at the exception of the
zeroth degree coefficient \eqref{eq:gamma0}. In the next section, we
turn to the computation of the coefficients of the degree $2$
for finite $N$, recovering the results found in
reference~\cite[Section 2.4]{Bertini:2007Stochastic}. 

\subsection{Degree-$2$ contributions: covariant
  matrix \label{sec:KMPdeg2}} 

Thus far, and thanks to the simplicity of equation
\eqref{eq:KMPbetanfromdt}, we have not had to cope with the
computation of coefficients $\gamma$ involving terms of degree larger
or equal to $2$ in the expansion of the stationary state
\eqref{eq:KMPness}. The arguments leading to equation
\eqref{eq:KMPbetanfromdt} are in fact more general and extend to
arbitrary order; see \ref{app:high}. The identification of the
degree-$2$ contributions to the stationary state \eqref{eq:KMPness}
occurs from the stationarity of the elements of the covariant matrix, 
\begin{equation}
  \label{eq:covmat}
  \beta_{m} \, \beta_{n} \avg{\xi_{m} \xi_{n}} - 1 - \delta_{m,n}
  =
    \begin{cases}
      \gamma_{m:1, n:1} \,, & m < n\,,\\  
      2 \gamma_{n:2} \,, & m = n\,,\\
      \gamma_{n:1, m:1}\,, & m > n\,,
    \end{cases}    
\end{equation}
with the boundary conditions $\avg{\xi_{\pm\sfrac{N}{2}\pm1} \xi_{n}}
= \beta_{\pm}^{-1} \beta_{n}^{-1}$ and
$\avg{\xi_{\pm\sfrac{N}{2}\pm1}^2} = 2\beta_{\pm}^{-2}$, i.e.,
\begin{subequations}
  \label{eq:KMPbetamnbc}
  \begin{align}
    \label{eq:KMPbetamnbcdiag}
     & \gamma_{\pm\sfrac{N}{2}\pm1:2} = 0\,,
    \\
    \label{eq:KMPbetamnbcoffdiag}
    \begin{split}
      &\gamma_{-\sfrac{N}{2}-1:1,n:1} = 0\,,\\
      &\gamma_{n:1,\sfrac{N}{2}+1:1} = 0\,,
    \end{split}
  \end{align}
\end{subequations}
where $-\sfrac{N}{2} \leq n \leq \sfrac{N}{2}$. 

Considering the stationarity of the quadratic energy moments
$\avg{\xi_{m} \xi_{n}} = 0\,$, $-\tfrac{N}{2} \leq m \leq n \leq
\tfrac{N}{2}$, we obtain results equivalent to those described in
reference~\cite{Bertini:2007Stochastic}:
\begin{enumerate}
\item
  for the diagonal elements,  
  \begin{multline}
    \label{eq:KMPbetan2}
    \beta_{n-1}^{-2} \gamma_{n-1:2} 
    + \beta_{n+1}^{-2} \gamma_{n+1:2} 
    - 4 \beta_{n}^{-2} \gamma_{n:2}
    + \beta_{n-1}^{-1} \beta_{n}^{-1} \gamma_{n-1:1,n:1} 
    + \beta_{n+1}^{-1} \beta_{n}^{-1} \gamma_{n:1,n+1:1} 
    \cr
    = -2
    \frac{(\beta_{+}^{-1} - \beta_{-}^{-1})^{2}}
    {(N+2)^{2}}
    \,;
  \end{multline}
  see equation \eqref{eq:pphigh_o2};
\item
  the off-diagonal elements, $m=n+1$, 
  \begin{multline}
    \label{eq:KMPbetannp1}
    \tfrac{1}{2}  (\beta_{n-1}^{-1} \beta_{n+1}^{-1} 
    \gamma_{n-1:1,n+1:1}
    + \beta_{n}^{-1} \beta_{n+2}^{-1} 
    \gamma_{n:1,n+2:1})
    - \tfrac{5}{3}  \beta_{n}^{-1} \beta_{n+1}^{-1} 
    \gamma_{n:1,n+1:1}
    \cr
    + \tfrac{1}{3} (\beta_{n}^{-2} \gamma_{n:2}
    + \beta_{n+1}^{-2}  \gamma_{n+1:2})
    = 
    \frac{1}{6}
    \frac{(\beta_{+}^{-1} - \beta_{-}^{-1})^{2}}
    {(N+2)^{2}}  
    \,,
  \end{multline}
  which is the combination of a contribution of degree $2$ from the
  pair $(n,n+1)$, as in equation \eqref{eq:pphigh_o2}, and two
  contributions of degree $1$ from the pairs $(n-1,n)$ and
  $(n+1,n+2)$, such as in \eqref{eq:pphigh_o1}; 
\item
  the off-diagonal elements with $|m-n|>1$ are the combinations of 
  contributions degree $1$ \eqref{eq:pphigh_o1}, 
  \begin{multline}
    \label{eq:KMPbetamn}
    \beta_{m-1}^{-1} \beta_{n}^{-1} 
    \gamma_{m-1:1,n:1} 
    + 
    \beta_{m}^{-1} \beta_{n-1}^{-1}
    \gamma_{m:1,n-1:1}
    + 
    \beta_{m+1}^{-1} \beta_{n}^{-1}
    \gamma_{m+1:1,n:1}
    \cr
    + 
    \beta_{m}^{-1} \beta_{n+1}^{-1} 
    \gamma_{m:1,n+1:1}
    - 4
    \beta_{m}^{-1} \beta_{n}^{-1} 
    \gamma_{m:1,n:1} 
    = 0
    \,.
  \end{multline}
\end{enumerate}
By symmetry of the elements of the covariant matrix, equations
\eqref{eq:KMPbetan2}-\eqref{eq:KMPbetamn} provide a closed set of
$N(N+1)/2$ equations for the coefficients $\gamma_{n:2}$ and
$\gamma_{m:1,n:1}$, with $m< n$, which must be solved for the inverse
temperatures \eqref{eq:KMPbetan} and boundary conditions
\eqref{eq:KMPbetamnbc}. 

To infer a solution of this set of equations, notice that equation
\eqref{eq:KMPbetamn} is actually a discrete form of the Poisson
equation and admits among its non-trivial solutions bilinear functions
of the form
\begin{equation}
  \label{eq:KMPbetamnsol}
  \beta_{m}^{-1} \beta_{n}^{-1} \gamma_{m:1,n:1}
  \propto
  \left(\frac{1}{2} + \frac{m}{N+2}\right)
  \left(\frac{1}{2} - \frac{n}{N+2}\right)
  \,,
\end{equation}
matching the boundary conditions \eqref{eq:KMPbetamnbcoffdiag}.
Remarkably, it is easy to identify a set of solutions of the system
\eqref{eq:KMPbetan2}-\eqref{eq:KMPbetamn} based on the form
\eqref{eq:KMPbetamnsol}. Matching this form to equations
\eqref{eq:KMPbetan2}-\eqref{eq:KMPbetannp1}, it is readily seen that 
bilinear solutions to equation \eqref{eq:KMPbetamn},
\begin{equation}
  \label{eq:KMPcovmat}
   \gamma_{m:1, n:1} 
  =
  \beta_{m} \, \beta_{n}
  \frac{(\beta_{+}^{-1} - \beta_{-}^{-1})^2}{N + 3}
  \left(\frac{1}{2} + \frac{m}{N+2} \right) 
  \left(\frac{1}{2} - \frac{n}{N+2} \right)\,,
\end{equation}
also solve equations \eqref{eq:KMPbetan2}-\eqref{eq:KMPbetannp1}, 
provided the diagonal elements of the covariant matrix are given by
\begin{equation}
  \label{eq:KMPcovmatdiag}
  \gamma_{n: 2} 
  = 
  \beta_{n}^{2}
  \frac{(\beta_{+}^{-1} - \beta_{-}^{-1})^2}{N + 3}
  \left[
    \frac{1}{4} - \frac{n^{2}}{(N+2)^{2}} + \frac{1}{2(N+2)}
  \right]
  \,.
\end{equation}
This solution, however, violates the boundary condition
\eqref{eq:KMPbetamnbcdiag}. To account for this correction, Fourier
modes must in general be added to equations
\eqref{eq:KMPcovmat}-\eqref{eq:KMPcovmatdiag},  
\begin{equation}
  \label{eq:KMPFouriermodes}
  \frac{1}{4(N+2)}
  \sum_{n_{1}, n_{2} = 1}^{2N+3}
  \big[ (-1)^{\tfrac{1}{2}(n_{1} + n_{2})}
  + (-1)^{\tfrac{3}{2}(n_{1} + n_{2})} \big]
  M_{n_{1}, n_{2}}
  \cos \pi ( n_{1} x + n_{2} y),
\end{equation}
with amplitudes $M_{n_{1}, n_{2}}$ which we will not explicitly
compute.

As noted by Bertini \emph{et al.} \cite{Bertini:2007Stochastic}, the
solution \eqref{eq:KMPcovmatdiag} becomes exact if alternatively one
changes the boundary condition  \eqref{eq:KMPbetamnbcdiag} to  
\begin{equation}
  \label{eq:KMPbetamnbcdiagalt}
  \gamma_{\pm\sfrac{N}{2}\pm1:2} = 
  \beta_{\pm}^2
  \frac{(\beta_{+}^{-1} - \beta_{-}^{-1})^2}{2(N + 2)(N + 3)}
  \,,
\end{equation}
which amounts to changing the energy distribution of the baths from
the canonical distribution \eqref{eq:canonicaldist} to 
\begin{equation}
  \label{eq:thermalbathalt}
  \beta_{\pm} \big[1 + \gamma_{\pm\sfrac{N}{2}\pm1:2}  
  L_2(\beta_{\pm} \xi) \big]\rme^{-\beta_{\pm} \xi} 
  \,.
\end{equation}  
Although this is not the canonical distribution expected of a thermal
bath, it can be interpreted as the distribution at the interface between
the thermal bath and the system. The temperature at the interface is
that of the thermal bath but a degree-$2$ correction proportional to
the square of the system's local  temperature gradient modifies its
energy distribution. Even then, correlations with the energy
distribution in the system are absent, as they should.

Pairwise correlations thus take a very simple form in the large
system-size limit, decaying with its inverse and falling off according
to a quadratic function of the locations of the two cells. 
These results
are similar to those discussed within an analogous framework in
reference~\cite{Nicolis:1984Onset}. They also arise in lattice gases
submitted to a density gradient \cite{Spohn:1983Long}.

\begin{figure}[htb]
  \centering
  \includegraphics[width=0.65\textwidth]
  {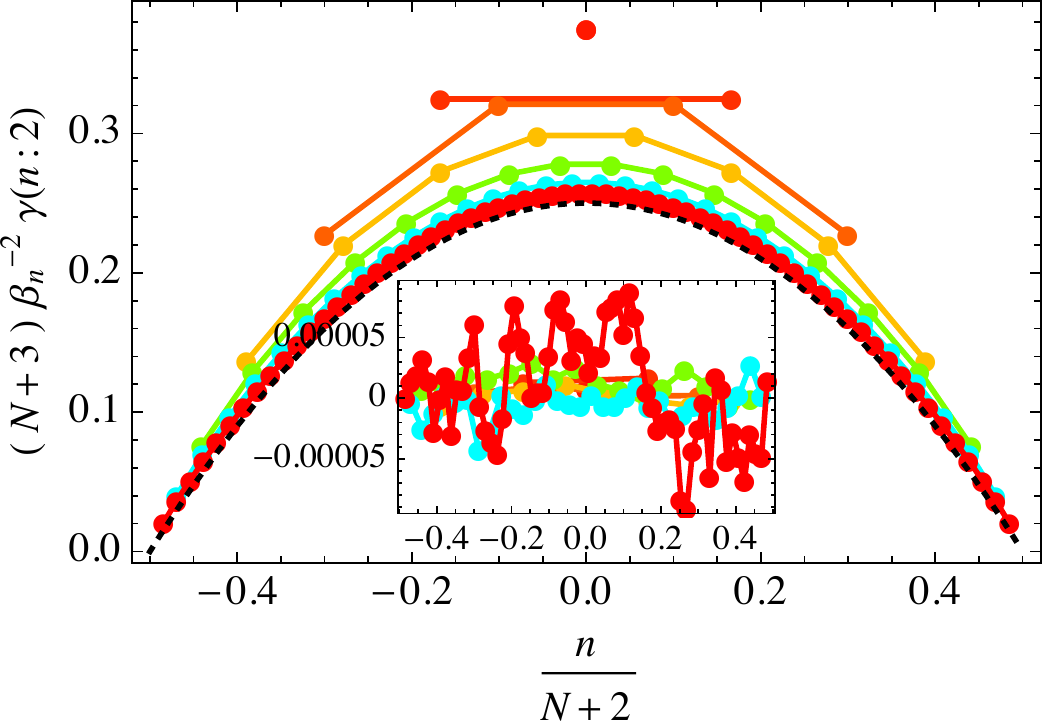}
  \caption{Computations of the diagonal elements of the covariant matrix
     $\gamma_{n:2}$ for system sizes $N+1 = 1, 2, \dots, 64$ under
     thermal boundary conditions \eqref{eq:thermalbathalt} with
     overall temperature difference $\beta_{+}^{-1} - \beta_{-}^{-1} = 1$. The
     black dotted curve shows the continuum limit $\tfrac{1}{4} - x^2$. The
     inset shows the unscaled differences between the analytic
     solution discussed in the text and numerical computations of
     these quantities. The error margin is controlled by the size of
     the sample.} 
  \label{fig:diagcovmat}
\end{figure}

\begin{figure}[htb]
  \centering
  \includegraphics[width=0.65\textwidth]{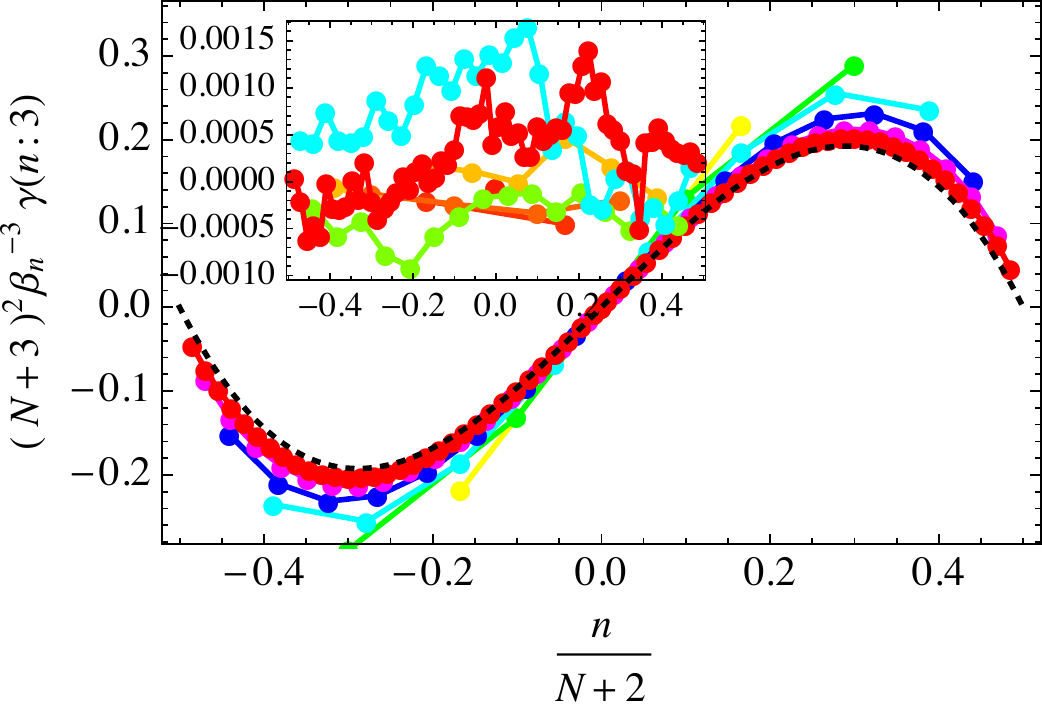}
  \caption{Similar to \fref{fig:diagcovmat} for the third-degree
    contributions $\gamma_{n:3}$. The black dotted curve is 
    $x(1 - 4 x^2)$.} 
  \label{fig:diagcovmat3}
\end{figure}

A numerical implementation of the KMP model subject to boundary
conditions \eqref{eq:thermalbathalt} is straightforward. A comparison
between  the diagonal elements $\gamma_{n:2}$ of the covariant matrix
\eqref{eq:KMPcovmatdiag} and numerical computations of these
quantities is shown in \fref{fig:diagcovmat} for systems of different
sizes ranging from  $N=0$ to $N=63$. The decay of  the
covariant matrix elements in direct proportion to the size of the
system, consistent with their disappearing in the infinite system-size
limit, as implied in reference~\cite{Kipnis:1982Heat}, makes them
difficult to measure accurately for large system sizes. 

As to higher-degree contributions to the non-equilibrium stationary
state, the results of \ref{app:high} allow their computation from sets
of closed linear algebraic equations which can in principle be carried
out for any system size. Simple analytic solutions similar to equations
\eqref{eq:KMPcovmat}-\eqref{eq:KMPcovmatdiag} are, however, more
difficult to obtain and it is not immediately clear whether conditions
exist on the energy distribution of the baths under which a
full-fledge expansion in terms of Fourier modes in the forme of equation
\eqref{eq:KMPFouriermodes} can be avoided. Their numerical computation
is also more difficult than second degree contributions due to their
faster decay; see \fref{fig:diagcovmat3}.

\subsection{Single-cell system \label{sec:KMPN0}}

For $N=0$, i.e.~a single cell interacting with two thermostats, it was
shown in reference~\cite{Bertini:2007Stochastic} that the stationary
state has energy distribution 
\begin{align}
  \label{eq:KMPnessN0}
  \PN{0} (\xi_{0}) 
  &= 
    \frac{\sqrt{\beta_{-}\,\beta_{+}}}{\pi} 
    \int_{\beta_{+}}^{\beta_{-}} \rmd \beta \, \rme^{-\beta\,\xi_{0}}
    \frac{1}{\sqrt{(\beta_{-} - \beta)(\beta - \beta_{+})}}\,,
    \cr
  &= 
    p_{\beta_{0}}(\xi_{0})
    \sum_{n_{0} = 0}^{\infty}  
    \frac{1 + (-)^{n_{0}}}{2^{n_{0}+1}}
    \binom{n_{0}}{\tfrac{n_{0}}{2}}
    \left(\frac{\beta_{-} - \beta_{+}} {\beta_{-} + \beta_{+}}\right)^{n_{0}}
    L_{n_{0}}(\beta_{0} \xi_{0}) 
  \,,
\end{align}
where the last line follows by expanding $\rme^{-\beta\,\xi_{0}}$
about $\rme^{-\beta_{0}\,\xi_{0}}$,  where $\beta_{0}^{-1} = \tfrac{1}{2}
(\beta_{+}^{-1} + \beta_{-}^{-1})$ is the cell's temperature, whose
value is prescribed by the requirement that $\PN{0}$ be
normalised. This last form is especially convenient for 
the sake of evaluating the moments of the distribution. 

We may use the results of \ref{app:high} to confirm the validity of
the solution \eqref{eq:KMPnessN0}. Considering the 
stationarity of the $n$\textsuperscript{th} energy moment, whose
contributions arise from equations
\eqref{eq:KMPLPleft}-\eqref{eq:KMPLPright}, the relevant terms are
obtained from equation \eqref{eq:apphigh_int2} with $c = n$ and
$d=0$. These terms multiply the coefficients $\gamma_{0:p}$  ($q=0$)
of the nonequilibrium stationary state \eqref{eq:KMPness}. Their sum
for $p = 0,\dots,n$, combined with minus twice the
$n$\textsuperscript{th} energy moment, must be equal to zero and
thus yields the value of $\gamma_{0:n}$ in terms of $\gamma_{0:p}$,
$p=0, \dots, n-1$ and parameter $\beta_{0}$,
\begin{equation}
  \label{eq:KMPnessN0sol}
  \gamma_{0:n} =
  \!
  \frac{(-)^{n}}{2n}
  \sum_{p=0}^{n-1} 
  (-)^{p} 
  \gamma_{0:p} 
  \left[
    \sum_{i = p}^{n-1} \beta_{0}^{n-i} 
    (\beta_{+}^{i-n} 
    +
    \beta_{-}^{i-n})
    \binom{i}{p}
    \!
    -
    2n
    \binom{n}{p} 
    \!    
  \right].  
\end{equation}
Letting $\gamma_{0:0} = 1$ and $\gamma_{0:1} = 0$, one first computes
$\beta_{0}$ from the $n=1$ term, then $\gamma_{0:n}$ for $n\geq2$. The
resulting series \eqref{eq:KMPnessN0} follows from the identity
\begin{equation}
  \label{eq:KMPnessN0ID}
    \sum_{p = 0}^{n-1}  
    \frac{1 + (-)^{p}}{2^{p+1}}
    \binom{p}{\tfrac{p}{2}}(1-x)^{p}
    \sum_{i=p}^{n-1} 
    \binom{i}{p}
    \left[x^{n-i} + (2-x)^{n-i}\right]
    =
    2n\, {}_{2}{F}_{1}(- \tfrac{n-1}{2},  -
    \tfrac{n}{2};1;(x-1)^{2})
    \,,
\end{equation}
where ${}_{2}{F}_{1}$ denotes the hypergeometric function \cite[\S
15.2]{WJOlver:2010p13722}. 

\subsection{Two-cell system \label{sec:KMPN1}}

The difficulty of extending the general result \eqref{eq:KMPnessN0} of
Bertini \emph{et al.} \cite{Bertini:2007Stochastic} to larger system
sizes is already apparent for $N=1$, i.e.~two cells, each of which is
in contact with a thermostat. Let $r 
\geq 0$ denote the degree. For each $n = 0, \dots, r$ and starting
from the bottom, the coefficients of degree $r$ are found by solving
the linear system of equations, derived through application of equation
\eqref{eq:apphigh_int2} to operators
\eqref{eq:KMPLP}-\eqref{eq:KMPLPright} acting on
$\xi_{\sfrac{-1}{2}}^{n} \, \xi_{\sfrac{1}{2}}^{r-n}$,
\begin{multline}
  \label{eq:KMPnessN1sol}
  \frac{1}{r+1}
  \sum_{p=0}^{r} \sum_{q = 0}^{r - p}
  (-)^{p+q}
  \gamma_{\sfrac{-1}{2}:p,\,\sfrac{1}{2}:q}
  \sum_{i = 0}^{r} 
  \binom{r - i}{p}
  \binom{i}{q}
  \beta_{\sfrac{-1}{2}}^{-(r - i)}
  \beta_{\sfrac{1}{2}}^{-i}
  \cr
  + 
  \frac{1}{n+1}
  \beta_{\sfrac{1}{2}}^{-(r - n)}
  \sum_{p = 0}^{n}
  \sum_{q = 0}^{r-n}
  (-)^{p+q}
  \binom{r-n}{q}
  \gamma_{\sfrac{-1}{2}:p,\,\sfrac{1}{2}:q}
  \sum_{i = 0}^{n} 
  \binom{n-i}{p}
  \beta_{\sfrac{-1}{2}}^{-(n - i)}
  \beta_{-}^{-i}
  \cr
  + 
  \frac{1}{r-n+1}
  \beta_{\sfrac{-1}{2}}^{-n}
  \sum_{p = 0}^{n}
  \sum_{q = 0}^{r-n}
  (-)^{p+q}
  \binom{n}{p}
  \gamma_{\sfrac{-1}{2}:p,\,\sfrac{1}{2}:q}
  \sum_{i = 0}^{r-n} 
  \binom{r-n-i}{q}
  \beta_{\sfrac{1}{2}}^{-(r - n-i)}
  \beta_{+}^{-i}
  \cr
  =
  3
  \beta_{\sfrac{-1}{2}}^{-n}
  \beta_{\sfrac{1}{2}}^{-(r-n)}
  \sum_{p = 0}^{n} 
  \sum_{q = 0}^{r-n} 
  (-)^{p+q}
  \binom{n}{p}
  \binom{r-n}{q}
  \gamma_{\sfrac{-1}{2}:p,\,\sfrac{1}{2}:q}  
\,.
\end{multline}

\begin{figure}[htb]
  \centering
  \includegraphics[width=0.65\textwidth]
  {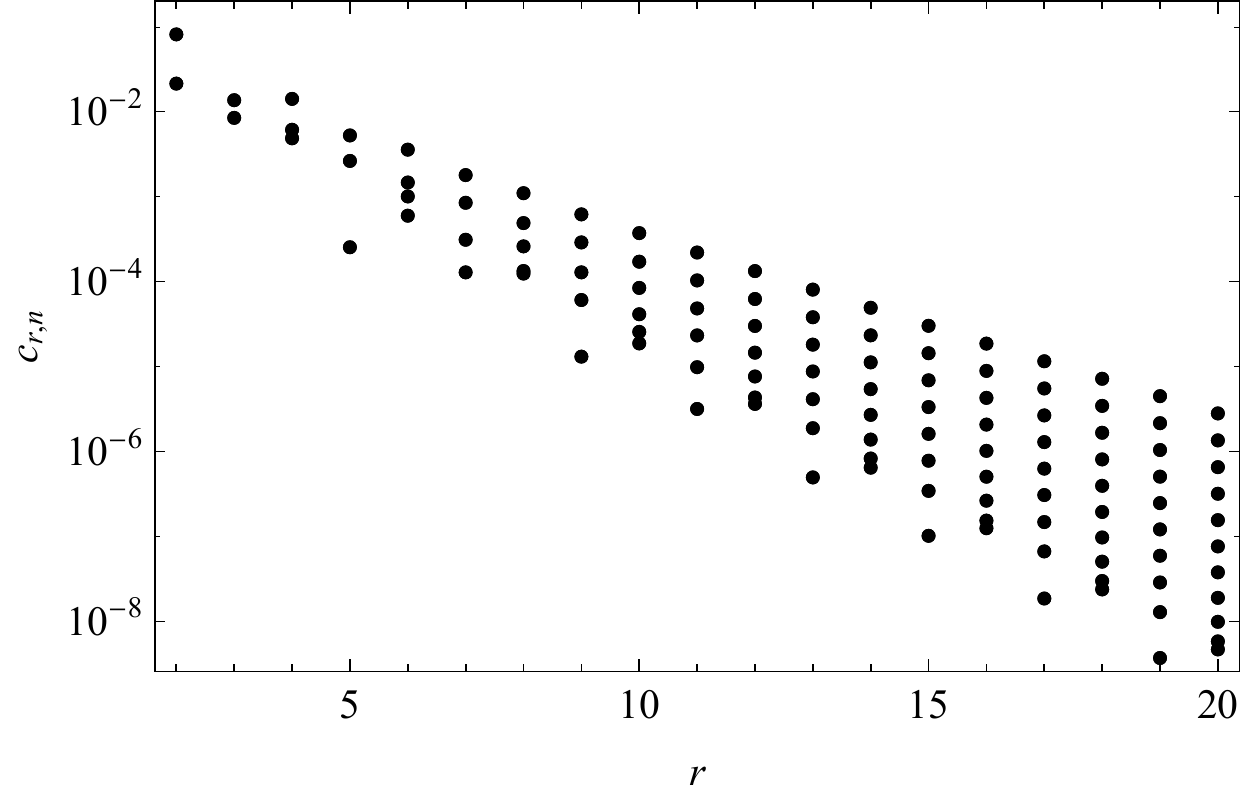}
  \caption{Coefficients $c_{r,n}$, equation
    \eqref{eq:KMPnessN1solgamma}, plotted as functions of their degree
    $r$, $2 \leq r \leq 20$. For each $r$, the values of $n$ decrease from
    $n=0$ to $n = \lfloor r/2 \rfloor$. For $r$ even,  the remaining
    values are symmetric with respect to $r/2$, $c_{r,r-n} = c_{r,n}$
    and, for $r$ odd, anti-symmetric, $c_{r,r-n} = -c_{r,n}$. Roughly
    speaking, the coefficients decay exponentially with $2^{-r}$.} 
  \label{fig:KMPnessN1sol}
\end{figure}

Letting $\gamma_{\sfrac{-1}{2}:0,\,\sfrac{1}{2}:0}   = 1$ and
$\gamma_{\sfrac{-1}{2}:1,\,\sfrac{1}{2}:0} =
\gamma_{\sfrac{-1}{2}:0,\,\sfrac{1}{2}:1}   = 0$, one first computes,
for $r=1$, 
\begin{equation}
  \label{eq:KMPnessN1solbeta}
  \begin{split}
    \beta_{\sfrac{-1}{2}}^{-1} & = \tfrac{1}{3}( 2 \beta_{-}^{-1} +
    \beta_{+}^{-1})\,,
    \\
    \beta_{\sfrac{1}{2}}^{-1} & = \tfrac{1}{3}( \beta_{-}^{-1} +
    2 \beta_{+}^{-1})\,.
  \end{split}
\end{equation}
Considering $r = 2,\,3,\dots$, one then finds, for $n = 0, \dots, \, r$,
the coefficients $\gamma_{\sfrac{-1}{2}:n,\,\sfrac{1}{2}:r -n}$. Letting
\begin{equation}
  \label{eq:KMPnessN1solgamma}
  c_{r,n} = 
  \frac{\beta_{\sfrac{-1}{2}}^{-n} \, \beta_{\sfrac{1}{2}}^{-(r-n)}}
  {( \beta_{+}^{-1} - \beta_{-}^{-1})^{r}}
  \gamma_{\sfrac{-1}{2}:n,\,\sfrac{1}{2}:r -n} \,,
\end{equation}
we obtain the results reported in table \ref{tab:KMPnessN1sol} and
\fref{fig:KMPnessN1sol} where values are displayed for $2 \leq r \leq 20$,
which involves $119$ different coefficients (up to a sign for $r$ odd).

\begin{table}[hbt]
  \begin{center}
    \begin{tabular}{|*{7}{c|}}
      \hline
      $r \backslash n$ &
      $0$ & $1$ & $2$ & $3$ & $4$ & $5$ \\
      \hline
      $2$ & $\frac{19}{234}$ & $\frac{5}{234}$ & 
            $\frac{19}{234}$ & & & \\
      $3$ & $\frac{67}{4914}$ & $\frac{77}{9126}$ &
            $\frac{-77}{9126}$ & $\frac{-67}{4914}$ & & \\
      $4$ & $\frac{16231}{1149876}$ & $\frac{1394}{287469}$ & 
            $\frac{781}{127764}$ & $\frac{1394}{287469}$ & 
            $\frac{16231}{1149876}$ & \\
      $5$ & $\frac{3824}{729729}$ & $\frac{104123}{39670722}$ 
         & $\frac{25271}{100039212}$ & $\frac{-25271}{100039212}$ 
         & $\frac{-104123}{39670722}$ & $\frac{-3824}{729729}$ 
      \\
      \hline
    \end{tabular}
  \end{center}
  \caption{Coefficients $c_{r,n}$, equation
    \eqref{eq:KMPnessN1solgamma}, for degrees $r=2,\dots,5$, obtained
    by solving equation \eqref{eq:KMPnessN1sol}.}
  \label{tab:KMPnessN1sol}
\end{table}

\section{Parameter-dependent KMP models \label{sec:KMPalpha}}

Generalisations of the KMP kernel \eqref{eq:KMPkernel} have been
obtained as instantaneous thermalisation limits of so-called Brownian
energy processes \cite{Giardina:2009Duality}. Here we consider the
kernels 
\begin{equation}
  \label{eq:KMPalphakernel}
  \Ka[\underline{\alpha}](\xi_{a}, \xi_{b}\rightarrow \xi'_{a}, \xi'_{b}) =
  \nu
  \frac{\Gamma(\alpha_{a} + \alpha_{b})}
  {\Gamma(\alpha_{a}) \Gamma(\alpha_{b})}
  \frac{{\xi'_{a}}^{\alpha_{a} - 1} \, {\xi'_{b}}^{\alpha_{b} - 1} }
  {(\xi_{a} + \xi_{b})^{\alpha_{a} + \alpha_{b} - 1}}
  \deltad( \xi_{a} + \xi_{b} - \xi'_{a} - \xi'_{b}) 
  \thetah(\xi'_{a}) \thetah(\xi'_{b}) \delta_{|a-b|,1}\,,
\end{equation}
where $\underline{\alpha} = \{\alpha_{-\sfrac{N}{2}}, \dots,
\alpha_{\sfrac{N}{2}}\}$ is a sequence of positive parameters,
$\alpha_{i} >0$, whose values may depend on the cell index. The KMP
kernel \eqref{eq:KMPkernel} is a particular case, retrieved for the
parameter values $\alpha_{i} = 1$, for all $i$. More generally, when
$\alpha_{i}$ has half-integer value, it is interpreted has half the
number of degrees of freedom involved in the local Brownian energy
process. Here we need not make such restrictions and consider
$\alpha_{i} \in \mathbb{R}_+$.

The kernels \eqref{eq:KMPalphakernel} share important properties of
the KMP kernel. They 
satisfy the detailed balance condition \eqref{eq:detailedbalance} with
equilibrium distribution now specified by Dirichlet distributions
which, for $N$ large, tend to the product of gamma distributions with
shape parameter $\alpha_{i}$ and scale parameter given 
by the temperature $\beta^{-1}$,  
\begin{equation}
  \label{eq:gammadist}
  p_{\alpha, \beta}( \xi) = \frac{\beta^{\alpha} \xi^{\alpha -
      1}}{\Gamma(\alpha)} \, \rme^{-\beta \, \xi}  
  \,.
\end{equation}
The canonical distribution \eqref{eq:canonicaldist} is recovered when
$\alpha = 1$. Furthermore, although the kernels exhibit  
explicit dependence on the outgoing energies $\xi'_{a}$ and
$\xi'_{b}$, their zeroth moment is actually uniform,
\begin{equation} 
  \label{eq:KMPalphafrequency}
  f_{\alpha_{1},\alpha_{2}} (\xi_1, \xi_2)  = \int \rmd \xi'_{1} \, \rmd\xi'_{2}\, 
  K_{\underline{\alpha}} (\xi_{1}, \xi_{2}\rightarrow \xi'_{1}, \xi'_{2}) = \nu \,,
\end{equation}
and consequently identical to equation
\eqref{eq:KMPfrequency}. Irrespective of the actual configuration
$\underline{\alpha}$ of shape parameters, all energy pairs thus
exchange energy at the same rate $\nu$. 

Moreover, the current, given by the first moment of the kernels
\eqref{eq:KMPalphakernel},  
\begin{equation} 
  \label{eq:KMPalphacurrent}
  j_{\alpha_{1},\alpha_{2}} (\xi_1, \xi_2)  = \int \rmd \xi'_{1} \,
  \rmd\xi'_{2}\, 
  (\xi_{1} - \xi'_{1})
  K_{\underline{\alpha}} (\xi_{1}, \xi_{2}\rightarrow \xi'_{1},
  \xi'_{2}) 
  = \nu \frac{\alpha_{1} \, \alpha_{2}} {\alpha_{1} + \alpha_{2}}
  \left(\frac{\xi_{1}}{\alpha_{1}}  - \frac{\xi_{2}}{\alpha_{2}}\right)
  \,,
\end{equation}
exhibits a dependence on the parameters whenever $\alpha_{1} \neq
\alpha_{2}$. Thus it might appear that, unless the factor $\alpha_{1}
\, \alpha_{2}/(\alpha_{1} + \alpha_{2})$ is independent of the pair of
cells through which the measurement is taking place, the current
\eqref{eq:KMPalphacurrent} is not expressible as the difference of a
local function evaluated at the two cells and does not  have  the
gradient property \cite{Spohn:1991book}. Nevertheless, more general
cases can be considered and their conductivity characterised, as we
shall see shortly.

By analogy with equation \eqref{eq:KMPness}, the non-equilibrium
stationary state $\PN{N}$ of this process can be expanded about the
local equilibria \eqref{eq:gammadist} in terms of polynomials, 
\begin{equation}
  \label{eq:KMPalphaness}
  \PN{N} (\xiN) = \prod_{n = \sfrac{-N}{2}}^{\sfrac{N}{2}} 
  p_{\alpha_{n}, \beta_n}(\xi_{n})
  \!\!\!\!\sum_{n_{\sfrac{-N}{2}}, \dots, n_{\sfrac{N}{2}} = 0}^{\infty}  
  \gamma_{n_{\sfrac{-N}{2}} \dots n_{\sfrac{N}{2}}}
  \Ja[\sfrac{-N}{2}]_{n_{\sfrac{-N}{2}}}(\beta_{\sfrac{-N}{2}} \xi_{\sfrac{-N}{2}}) \dots
  \Ja[\sfrac{N}{2}]_{n_{\sfrac{N}{2}}}(\beta_{\sfrac{N}{2}} \xi_{\sfrac{N}{2}})
  \,,
\end{equation}
where the polynomials $\Ja_{n}$, 
\begin{equation}
  \label{eq:Ja}
  \Ja_{n} (x) = 
  \sqrt{\frac{\Gamma(\alpha) \, n!}
    {\Gamma(n + \alpha)}}
  L^{\alpha-1}_{n}(x)
\end{equation}
define a complete set derived from the generalised Laguerre
polynomials of parameter $\alpha - 1$, whose elements are 
orthonormal with respect to the weight function $p_{\alpha,1}(x)$,
as specified by equation \eqref{eq:gammadist}.

\subsection{Heat current and temperature profile \label{sec:KMPalphacurrent}}

Taking the average of the current \eqref{eq:KMPalphacurrent} with
respect to the stationary state \eqref{eq:KMPalphaness}, we have
\begin{equation}
  \label{eq:KMPalphaFourierlocal}
  \avg{ j(\xi_n, \xi_{n+1}) } = - \nu 
  \frac{\alpha_{n} \, \alpha_{n+1}} 
  {\alpha_{n} + \alpha_{n+1}}
  (\beta_{n+1}^{-1} -
  \beta_{n}^{-1})\,.
\end{equation}
Considering the total current, $\Jh(N+2) = \sum_{n} \avg{ j(\xi_n, \xi_{n+1})
}$, since we must have $\avg{ j(\xi_n, \xi_{n+1}) } = \Jh(N+2)/(N+2)$,
we find
\begin{equation}
  \label{eq:KMPalphaFourierglobal}
  \avg{ j(\xi_n, \xi_{n+1}) }   
  = - \frac{\nu}{2} \frac{\beta_{+}^{-1} - \beta_{-}^{-1}}{N+2}
  \left[ \frac{\sum_{n = \sfrac{-N}{2}}^{\sfrac{N}{2}} \alpha_{n}^{-1} 
      + \tfrac{1}{2}(\alpha_{-}^{-1} + \alpha_{+}^{-1})}{N+2} \right]^{-1}
  \,.
\end{equation}
Letting $N\to\infty$, provided the overall temperature gradient of the
nonequilibrium stationary state is linear, which may depend on the 
configuration of shape parameters $\underline{\alpha}$, the heat
conductivity is given by $\nu/2$ times the harmonic mean of the shape
parameters,
\begin{equation}
  \label{eq:KMPalphakappa}
  \kappa_{\underline{\alpha}} 
  = \frac{\nu}{2} \langle \underline{\alpha}^{-1} \rangle^{-1}\,.
\end{equation}
This is remindful of the problem of diffusion in one-dimensional
disordered lattices \cite{Bernasconi:1978Classical}.

To determine the local temperatures and verify the linearity of the
overall temperature profile, we note that the results obtained 
for the characterisation of the non-equilibrium stationary state of
the KMP model transpose to the kernels \eqref{eq:KMPalphakernel}; see 
\ref{app:alpha}. In particular, considering the stationarity of the
first energy moments, we obtain the following identity governing the
temperature profile, 
\begin{equation}
  \label{eq:KMPalphabetan}
  \beta_{n}^{-1} [\alpha_{n+1} (\alpha_{n-1} + \alpha_{n}) +
  \alpha_{n-1} (\alpha_{n+1} + \alpha_{n})]   
  = \beta_{n+1}^{-1} \alpha_{n+1} (\alpha_{n-1} + \alpha_{n})  + 
  \beta_{n-1}^{-1} \alpha_{n-1} (\alpha_{n+1} + \alpha_{n})  
  \,.
\end{equation}

There are two simple cases such that this equation reduces to
\eqref{eq:KMPbetan}: 
\begin{enumerate}
\item \label{case:1}
  if all $\alpha_{n}$ are identical, or
\item \label{case:2}
  if they alternate between two different values, say $\alpha_{0}$ and
  $\alpha_{1}$, depending on the parity of $n$.
  \setcounter{tempcounter}{\value{enumi}}
\end{enumerate}
In both cases, the temperature profile is strictly linear and given by
\eqref{eq:KMPbetanN}. The current \eqref{eq:KMPalphacurrent} has the
gradient property and the heat conductivity of the resulting process
\eqref{eq:KMPalphakappa} can also be inferred in terms of the static
correlations only,
\begin{equation}
  \label{eq:KMPalphastatic}
  \kappa_{\alpha_{1},\alpha{2}} = 
  \frac{\beta^{2}}{2} \int_{0}^{\infty} \rmd\xi_{1} \, \rmd\xi_{2}
  (\xi_{1} - \xi_{2}) j_{\alpha_{1},\alpha_{2}} (\xi_1, \xi_2)
  p_{\alpha_{1}, \beta}( \xi_{1})
  p_{\alpha_{2}, \beta}( \xi_{2})
  = \nu
  \frac{\alpha_{1} \, \alpha_{2}}
  {\alpha_{1}  + \alpha_{2}}
  \,.
\end{equation}
Perhaps interestingly, this value is equal to the exchange frequency
\eqref{eq:KMPalphafrequency} when $\alpha_{1} \alpha_{2} = \alpha_{1}
+ \alpha_{2}$, which has positive solutions 
for $\alpha_{1}, \, \alpha_{2} >1$. Such solutions include, in
particular, $\alpha_{1} = \alpha_{2} = 2$ and $\alpha_{1} =
\tfrac{3}{2}$, $\alpha_{2} = 3$, for which this identity may therefore
be given a mechanical  interpretation. 

By extension, a linear temperature profile is
also observed   
\begin{enumerate}
  \setcounter{enumi}{\value{tempcounter}}
\item \label{case:3}
  if the sequence of shape parameters is periodic, i.e.~such that
  $\alpha_{n} = \alpha_{n\,\mathrm{mod}\,p}$ where $p\geq 3$ is the
  period.    
  \setcounter{tempcounter}{\value{enumi}}
\end{enumerate}
However, in such cases, and assuming $N+2$ is a multiple of $p$,
periodic excursions of lengths $p$ are observed about the linear
temperature profile \eqref{eq:KMPbetanN}, 
\begin{equation}
  \label{eq:KMPbetanNalphap}
  \beta_{n}^{-1} = \tfrac{1}{2} ( \beta_{+}^{-1} + \beta_{-}^{-1} ) 
  + \frac{n + \theta_{n\,\mathrm{mod}\,p}}{N+2} ( \beta_{+}^{-1} - \beta_{-}^{-1} ) 
  \,.
\end{equation}
Letting $\gamma^{(+)}_{n} = \alpha_{n+1} ( \alpha_{n} + \alpha_{n-1}
)$ and $\gamma^{(-)}_{n} = \alpha_{n-1} ( \alpha_{n} + \alpha_{n+1}
)$, with $n = 0,\dots,p-1$, the coefficients $\theta_{0}$, \dots,
$\theta_{p-1}$ are determined through the set of $p$ equations,
\begin{equation}
  \label{eq:KMPbetanNalphaptheta}
  \gamma^{(+)}_{n} 
  \left(\theta_{n+1\,\mathrm{mod}\,p} - \theta_{n}+ 1 \right)  
  + \gamma^{(-)}_{n} 
  \left( \theta_{n-1\,\mathrm{mod}\,p} - \theta_{n}- 1 \right)
  = 0
  \,.
\end{equation}

Other cases of interest are those of a random sequences of shape
parameters, for which the temperature profiles have the form
\eqref{eq:KMPbetanNalphap}-\eqref{eq:KMPbetanNalphaptheta}, but
without the periodicity. Among such models, different classes can be
distinguished, in particular: 
\begin{enumerate}
  \setcounter{enumi}{\value{tempcounter}}
\item \label{case:4}
  if the shape parameters are drawn randomly from a finite set,
\item \label{case:5}
  if the shape parameters are drawn randomly from the (countable) set
  of positive half integers, or
\item \label{case:6}
  if the shape parameters are drawn randomly over the real positive numbers.
  \setcounter{tempcounter}{\value{enumi}}
\end{enumerate}
The last category \eqref{case:6} is a priori problematic, in particular
when the values of the shape parameters can be arbitrarily
small. Loosely speaking, in such cases, the temperature profile
typically exhibits a step-like structure with points of discontinuity
at sites where the shape parameters are small. Such a situation is
excluded in cases \eqref{case:4}-\eqref{case:5}. In case
\eqref{case:4}, fluctuations about the linear temperature profile are
expected to be finite. Case \eqref{case:5} is interesting since, in the
framework of Brownian energy processes, it amounts to randomly
selecting the number of degrees of freedom at every site (with respect
to some probability distribution on the set). Generally
speaking, one expects equation \eqref{eq:KMPalphakappa} to hold
provided a linear temperature profile is recovered upon local
averaging at some intermediate scale, that is, provided $\theta_{n}$
vanishes when locally averaged on this intermediate scale.

We will not dwell further on such considerations which touch upon
the broader problem of random walks in random environments
\cite{Solomon:1975Random, Sinai:1983limiting} and bears similarities
with the problem of conduction in random one-dimensional chains
\cite{Alexander:1981Excitation}, as well as,
in higher dimensions, the percolation threshold
\cite{Kirkpatrick:1973Percolation}; it  deserves a separate
study. Rather, we focus below on models  \eqref{case:1} and 
\eqref{case:2} and obtain the second-degree contributions to their 
non-equilibrium stationary states, from which the contributions to the
covariant matrix are deduced. 

\subsection{Uniform $\alpha$ \label{sec:KMPalphauniform}}

Letting $\alpha_{i} \equiv \alpha$, we have the temperature profile
\eqref{eq:KMPbetanN} and stationary current,
\begin{equation}
  \label{eq:KMPalphauniformFourier}
  \avg{ j(\xi_n, \xi_{n+1}) } = - 
  \kappa_{\alpha}
  \frac{\beta_{+}^{-1} - \beta_{-}^{-1}}
  {N+2}\,,
\end{equation}
with heat conductivity
\begin{equation}
  \label{eq:KMPalphauniformkappa}
  \kappa_{\alpha} =   \frac{\nu \, \alpha}{2}\,.
\end{equation}

Considering the second degree contributions to the nonequilibrium
stationary state, one finds that equation \eqref{eq:KMPbetamn} remains
unchanged. To find the second degree contributions, we observe that 
equations \eqref{eq:KMPbetan2}-\eqref{eq:KMPbetannp1} transpose to:
\begin{multline}
  \label{eq:KMPalphauniformbetan2}
  (\alpha + 1)
  (\beta_{n-1}^{-2} \gamma_{n-1:2} 
  + \beta_{n+1}^{-2} \gamma_{n+1:2})
  + 
  \sqrt{2\alpha(\alpha + 1)}
  (\beta_{n-1}^{-1} \beta_{n}^{-1} \gamma_{n-1:1,n:1} 
  + \beta_{n+1}^{-1} \beta_{n}^{-1} \gamma_{n:1,n+1:1})
  \cr
  -2 (3\alpha + 1) 
  \beta_{n}^{-2} \gamma_{n:2}
  = -\sqrt{2\alpha(\alpha+1)^{3}}
  \frac{(\beta_{+}^{-1} - \beta_{-}^{-1})^{2}}
  {(N+2)^{2}}
  \,,
\end{multline}
and
\begin{multline}
  \label{eq:KMPalphauniformbetannp1}
  (2\alpha + 1)  
  (\beta_{n-1}^{-1} \beta_{n+1}^{-1} 
  \gamma_{n-1:1,n+1:1}
  + \beta_{n}^{-1} \beta_{n+2}^{-1} 
  \gamma_{n:1,n+2:1})
  - 2(3\alpha + 2)  \beta_{n}^{-1} \beta_{n+1}^{-1} 
  \gamma_{n:1,n+1:1}
  \cr
  + \sqrt{2\alpha(\alpha + 1)}
  (\beta_{n}^{-2} \gamma_{n:2}
  + \beta_{n+1}^{-2} \gamma_{n+1:2})
  = 
  \alpha^{2}
  \frac{(\beta_{+}^{-1} - \beta_{-}^{-1})^{2}}
  {(N+2)^{2}}  
  \,.
\end{multline}
Equations \eqref{eq:KMPcovmat}-\eqref{eq:KMPcovmatdiag} thus become 
to 
\begin{equation}
  \label{eq:KMPalphauniformcovmat}  
   \gamma_{m: 1, n: 1} 
  =
  \beta_{m} \, \beta_{n}
  \frac{(\beta_{+}^{-1} - \beta_{-}^{-1})^2}
  {N + 2 + \alpha^{-1}}
  \left(\frac{1}{2} + \frac{m}{N+2}\right)
  \left(\frac{1}{2} - \frac{n}{N+2}\right)
  \,,
\end{equation}
and
\begin{equation}
  \label{eq:KMPalphauniformcovmatdiag}
  \gamma_{n: 2} 
  = 
  \beta_{n}^{2} 
  \sqrt{\frac{\alpha + 1}{2 \alpha}}
  \frac{(\beta_{+}^{-1} - \beta_{-}^{-1})^2}
  {N + 2 + \alpha^{-1}}
  \left[
    \frac{1}{4} - \frac{n^{2}}{(N+2)^{2}} + \frac{\alpha}{2 (N+2)}
  \right]
  \,.
\end{equation}

As with $\alpha = 1$, the boundary condition
\eqref{eq:KMPbetamnbcdiag} has to be modified for equation
\eqref{eq:KMPalphauniformcovmatdiag} to become an exact solution. We
must therefore impose
\begin{equation}
  \label{eq:KMPalphauniformbetamnbcdiagalt}
  \gamma_{\pm\sfrac{N}{2}\pm1:2} = 
  \beta_{\pm}^2
  \sqrt{\tfrac{1}{2}\alpha(\alpha + 1)}
  \frac{(\beta_{+}^{-1} - \beta_{-}^{-1})^2}
  {(N + 2 + \alpha^{-1}) (N + 2)}
  \,,
\end{equation}
which amounts to changing the energy distribution of the baths to 
\begin{equation}
  \label{eq:KMPalphauniformthermalbathalt}
  \big[1 + \gamma_{\pm\sfrac{N}{2}\pm1:2}  
  \Ja_{2}(\beta_{\pm} \xi) \big]
  p_{\alpha, \beta_{\pm}}(\xi)
  \,.
\end{equation}  

The limit $\alpha\to0$ is of particular interest. The integrated
kernel \eqref{eq:KMPalphakernel} corresponding to this regime,  
\begin{equation}
  \label{eq:KMPalphakernelint}
  \lim_{\alpha\to0}
  \int_{-\xi_{b}}^{\eta} \rmd h
  \, \Ka[\alpha](\xi_{a}, \xi_{b} \to \xi_{a} - h, \xi_{b} + h)
  = 
  \lim_{\alpha\to0}
  \nu
  \frac{\Gamma(2\alpha)}
  {\Gamma(\alpha)^{2}}
  \int_{0}^{(\eta+\xi_{b})/(\xi_{a} + \xi_{b})} \rmd x
  (1 - x)^{\alpha - 1} \, x^{\alpha - 1} \,,
\end{equation}
is the cumulative distribution function of a regularised Beta
distribution with vanishing shape parameter. Its value thus tends to
$1/2$ for $-\xi_{b} < \eta < \xi_{a}$. The limiting energy exchange
process thus induces a complete transfer of energy to either of the
two interacting cells, $\eta = -\xi_{b}$ or $\eta = \xi_{a}$, with
probability rate $\nu/2$, viz.
\begin{equation}
  \label{eq:KMPalphakernela->0}
  \Ka[0](\xi_{a}, \xi_{b}\rightarrow \xi'_{a}, \xi'_{b}) =  
  \frac{\nu}{2} 
  \Big[
  \deltad( \xi_{a} + \xi_{b} - \xi'_{a}) \deltad(\xi'_{b}) 
  +
  \deltad( \xi_{a} + \xi_{b} - \xi'_{b}) \deltad(\xi'_{a}) 
  \Big] \delta_{|a-b|,1}\,.
\end{equation}
In other words $p$ in equation \eqref{eq:xipxi}
takes its values in the discrete set $\{0,1\}$, each with probability
$1/2$. In an empty neighbourhood, energy packets thus perform a random
walk, coalescing whenever two energy packets overlap.

The fact that the heat conductivity \eqref{eq:KMPalphauniformkappa}
goes to zero in this limit is a feature of the anomalously slow
kinetics of heat transfer which results from the coupling of diffusion
and aggregation \cite{BenAvraham:1990Statics}. As a simple
illustration, let  $\alpha \equiv \alpha_{N}$,
\begin{equation}
  \label{eq:KMPalphauniformalphaNto0}
  \alpha_{N} = \frac{2 \alpha_{0}}{N+2}\,,
\end{equation}
with $\alpha_{0} >0$, in which case the stationary current
\eqref{eq:KMPalphauniformFourier} scales with $(N+2)^{-2}$. 

With this choice of parametrisation, Equations
\eqref{eq:KMPalphauniformcovmat}-\eqref{eq:KMPalphauniformcovmatdiag} 
yield respectively 
\begin{equation}
  \label{eq:KMPalphauniformcovmata->0}
  \gamma_{m:1,n:1} = 
  \beta_{m} \, \beta_{n} \frac{2 \alpha_{0}}{1+2 \alpha_{0}}
  \frac{(\beta_{+}^{-1} - \beta_{-}^{-1})^2}{N+2}
  \left(\frac{1}{2} + \frac{m}{N+2}\right)
  \left(\frac{1}{2} - \frac{n}{N+2}\right)\,,
\end{equation}
and
\begin{equation}
  \label{eq:KMPalphauniformcovmatdiaga->0}
  \gamma_{n:2} = 
  \beta_{n}^{2} \sqrt{\frac{\alpha_{0}}{N+2} 
    \left( 1 + \frac{2 \alpha_{0}}{N+2} \right) } 
  \frac{(\beta_{+}^{-1} - \beta_{-}^{-1})^2}{1 + 2 \alpha_{0}}
  \left[
    \frac{1}{4} - \frac{n^{2}}{(N+2)^{2}}
    + \frac{\alpha_{0}}{(N+2)^{2}}
  \right]\,.
\end{equation}
The off-diagonal elements thus behave similarly to equation
\eqref{eq:KMPalphauniformcovmat}, but the diagonal elements decay
with the square root of the system size, slower than equation
\eqref{eq:KMPalphauniformcovmatdiag}. At the same time, the
distribution \eqref{eq:KMPalphauniformthermalbathalt} 
converges to the usual gamma distribution faster than the coefficient
\eqref{eq:KMPalphauniformbetamnbcdiagalt} when $\alpha$ is fixed. 

The opposite limit, $\alpha\to\infty$, is such that the sum of the
energies of the interacting cells is exactly halved among them,
\begin{equation}
  \label{eq:KMPalphakernela->infty}
  \Ka[\infty](\xi_{a}, \xi_{b}\rightarrow \xi'_{a}, \xi'_{b}) =  
  \nu
  \deltad[ \tfrac{1}{2}(\xi_{a} + \xi_{b}) - \xi'_{a}]
  \deltad( \xi'_{a} - \xi'_{b})
  \delta_{|a-b|,1}\,.
\end{equation}
The sources of randomness are thus restricted to the interaction
times. As of energy transfers from the thermal boundaries, notice 
that, for large shape parameters, the energy distributions of the
baths are sharply peaked about $\alpha \, \beta_{\pm}^{-1}$. 
Energies at the boundaries are therefore fixed and, in the bulk,
fluctuations of the energies rescaled by the shape parameter,
$\alpha^{-1}\xi_{n}$, about the local temperature $\beta_{n}^{-1}$ are
inversely proportional to the system size. 

Correlations in this regime can be studied, for instance,
by assuming 
\begin{equation}
  \label{eq:KMPalphauniformalphaNtoinfty}
  \alpha_{N} = \tfrac{1}{2} \alpha_{0}(N+2)\,,
\end{equation}
for which the current \eqref{eq:KMPalphauniformFourier} is independent
of $N$ and, similarly to conduction in harmonic chains
\cite{Casher:1971Heat}, the heat conductivity
\eqref{eq:KMPalphauniformkappa} is infinite. 

Whereas the off-diagonal elements of the second-degree
contributions to the stationary state \eqref{eq:KMPalphauniformcovmat}
become independent of $\alpha_{0}$, the diagonal elements
\eqref{eq:KMPalphauniformcovmatdiag} still decay with $(N+2)^{-1}$ but
pick up a uniform contribution proportional to $\alpha_{0}$, 
\begin{equation}
  \label{eq:KMPalphauniformcovmatdiaga->inf}
  \gamma_{n: 2} 
  \approx 
  \beta_{n}^{2} 
  \frac{(\beta_{+}^{-1} - \beta_{-}^{-1})^2}
  {\sqrt{2}(N+2)}
  \left[
    \frac{1}{4} - \frac{n^{2}}{(N+2)^{2}} + \frac{\alpha_{0}}{4}
  \right]
  \,.
\end{equation}
 The second-degree contribution to the energy distribution of the
 baths \eqref{eq:KMPalphauniformbetamnbcdiagalt} must therefore scale
 with $N^{-1}$ rather than $N^{-2}$ for fixed $\alpha$.

\subsection{Alternating $\alpha_{0}$ and $\alpha_{1}$ 
\label{sec:KMPalpha01}}

For the sake of the argument, let us think of $N$ as even. Let
$i\in\{-\tfrac{N}{2},\dots,\tfrac{N}{2}\}$ and assume $\alpha_{i} =
\alpha_{0}$ if $i$ is even and $\alpha_{i} = \alpha_{1}$ if $i$ is odd
and denote by $\sigma(i) \in \{0,\,1\}$ the parity of $i$\footnote{For
  $N$ odd, the parity of $i$ should be interpreted as that of $i +
  \tfrac{1}{2}$.} , 
\begin{equation}
  \sigma(i)
  = 
  \begin{cases}
    0,& i\,\mathrm{even},\\
    1,& i\,\mathrm{odd}.
  \end{cases}
\end{equation}
Equations \eqref{eq:KMPbetan2}-\eqref{eq:KMPbetamn}
take on the expressions, for $m=n$,
\begin{multline}
  \label{eq:KMPalpha01betan2}
  \sqrt{\frac{1 + \alpha_{1 - \sigma(n)}}
    {\alpha_{1 - \sigma(n)}}}
  (\beta_{n-1}^{-2}\gamma_{n-1:2}   
  + \beta_{n+1}^{-2}\gamma_{n+1:2})
  \cr
  +
  \sqrt{2\frac{\alpha_{\sigma(n)}}
    {\alpha_{1-\sigma(n)}}}
  (\beta_{n-1}^{-1} \beta_{n}^{-1} 
  \gamma_{n-1:1,n:1} 
  +
  \beta_{n}^{-1} \beta_{n+1}^{-1} 
  \gamma_{n:1,n+1:1})
  \cr
  -
  2 \frac{1 + \alpha_{0} + \alpha_{1} + \alpha_{\sigma(n)}}
  {\sqrt{\alpha_{\sigma(n)}(1 + \alpha_{\sigma(n)})}}
  \beta_{n}^{-2}\gamma_{n:2}   
  = 
  -\sqrt{2}(1 + \alpha_{1-\sigma(n)})
  \frac{(\beta_{+}^{-1} - \beta_{-}^{-1})^{2}}
  {(N+2)^{2}}
  \,,
\end{multline}
for $m=n+1$,
\begin{multline}
  \label{eq:KMPalpha01betannp1}
  (1 + \alpha_{0} + \alpha_{1})
  (\beta_{n-1}^{-1} \beta_{n+1}^{-1} 
  \gamma_{n-1:1,n+1:1} 
  +
  \beta_{n}^{-1} \beta_{n+2}^{-1} 
  \gamma_{n:1,n+2:1})
  \cr
  +
  \sqrt{2 \alpha_{\sigma(n)} ( 1 + \alpha_{\sigma(n)} )}
  \beta_{n}^{-2} \gamma_{n:2}
  +
  \sqrt{2 \alpha_{\sigma(n+1)} ( 1 + \alpha_{\sigma(n+1)} )}
  \beta_{n+1}^{-2}\gamma_{n+1:2}   
  \cr
  -2 \frac{\alpha_{0} ( 1 + \alpha_{0} ) 
    + \alpha_{1} ( 1 + \alpha_{1} ) 
    + \alpha_{0} \alpha_{1} }
  {\sqrt{\alpha_{0} \alpha_{1}}}
  \beta_{n}^{-1} \beta_{n+1}^{-1} 
  \gamma_{n:1,n+1:1} 
  = 
  \alpha_{0} \alpha_{1}
  \frac{(\beta_{+}^{-1} - \beta_{-}^{-1})^{2}}
  {(N+2)^{2}}\,,
\end{multline}
and, for $|m-n|>1$,
\begin{multline}
  \label{eq:KMPalpha01betamn}
    \beta_{m-1}^{-1} \beta_{n}^{-1} 
    \gamma_{m-1:1,n:1} 
    + 
    \beta_{m}^{-1} \beta_{n-1}^{-1}
    \gamma_{m:1,n-1:1}
    + 
    \beta_{m+1}^{-1} \beta_{n}^{-1}
    \gamma_{m+1:1,n:1}
    + 
    \beta_{m}^{-1} \beta_{n+1}^{-1} 
    \gamma_{m:1,n+1:1}
    \cr
    = 2 \frac{\alpha_{1-\sigma(m)} + \alpha_{1-\sigma(n)}}
    {\sqrt{\alpha_{0}\, \alpha_{1}}}
    \beta_{m}^{-1} \beta_{n}^{-1} 
    \gamma_{m:1,n:1} 
\end{multline}

The solutions to equation \eqref{eq:KMPalpha01betamn} which match the
boundary conditions \eqref{eq:KMPbetamnbcoffdiag} take forms similar 
to that of equation \eqref{eq:KMPbetamnsol}, but with an extra factor
$\sqrt{\alpha_{0}/\alpha_{1}}$ if both $m$ and $n$ are even, or
$\sqrt{\alpha_{1}/\alpha_{0}}$ if both $m$ and $n$ are odd. By
matching them to equations \eqref{eq:KMPalpha01betan2} and
\eqref{eq:KMPalpha01betannp1}, one obtains the expressions:
\begin{equation}
  \label{eq:KMPalpha01betamnsol}
  \gamma_{m:1,n:1}
  = 2 
  \frac{\alpha_{0} \alpha_{1}}{
    \sqrt{\alpha_{\sigma(m)} \alpha_{\sigma(n)}}}
  \beta_{m} \, \beta_{n} 
  \frac{(\beta_{+}^{-1} - \beta_{-}^{-1})^{2}}
  {(N+2)(\alpha_{0} + \alpha_{1}) + 2}
  \left(\frac{1}{2} + \frac{m}{N+2}\right)
  \left(\frac{1}{2} - \frac{n}{N+2}\right)
  \,,
\end{equation}
and
\begin{equation}
  \label{eq:KMPalpha01betan2sol}
  \gamma_{n:2}
  = 
  \sqrt{2\alpha_{1-\sigma(n)} (1 + \alpha_{1-\sigma(n)})}
  \beta_{n}^{2} 
  \frac{(\beta_{+}^{-1} - \beta_{-}^{-1})^{2}}
  {(N+2)(\alpha_{0} + \alpha_{1})+2}
  \left[\frac{1}{4} - \frac{n^{2}}{(N+2)^{2}} + 
  \frac{\alpha_{\sigma(n)}}{2(N+2)}\right]
  \,.
\end{equation}
These solutions are exact provided the thermal baths have energy
distributions 
\begin{equation}
  \label{eq:KMPalpha01thermalbathalt}
  \big[1 + \gamma_{\pm\sfrac{N}{2}\pm1:2}  
  \Ja[\sigma(\pm)]_{2}(\beta_{\pm} \xi) \big]
  p_{\alpha_{\sigma(\pm)}, \beta_{\pm}}(\xi)
  \,.
\end{equation}  

\section{Concluding remarks \label{sec:con}}

The Kipnis--Marchioro--Presutti  model of heat conduction
\cite{Kipnis:1982Heat} belongs to a larger class of stochastic energy
exchange Markov jump processes derived from the instantaneous
thermalisation limit of Brownian energy processes
\cite{Giardina:2009Duality}. On the one hand, the defining common
feature of these models is that every pair of neighbouring cells
exchanges energy among them at uniform rate. Their distinctive
feature, on the other hand, is that the detailed balance condition is
obeyed with respect to different canonical equilibrium energy
distributions, which are identified by the sequence of shape
parameters associated with every model.

The non-equilibrium stationary states resulting from the application
of a temperature gradient at the system's boundaries are amenable to
analytic treatment. Indeed, the simple structure of the stochastic
kernel allows for the determination of the stationary states in terms
of the products of local canonical equilibrium distributions identified
by their shape parameters and a multinomial expansion specified by the 
orthogonal polynomials associated with them. The collection of all
coefficients of degree $r$ in this expansion is obtained by solving
sets of linear equations derived by invoking the stationarity of
$r$-point correlation functions. 

The temperature profile and covariant matrix elements are
thus obtained by considering the first and second degree terms in this
expansion. Among the models we considered, those specified by
sequences of alternating shape parameters yield a linear temperature
profile under thermal boundary conditions at different
temperatures. Explicit expressions of their covariant matrix elements
were obtained upon the condition that the energy distributions
associated with the thermal baths have second degree contributions
tailored so as to eliminate Fourier components of non-zero
wavelengths. This provides a generalisation of the results obtained by
Bertini \emph{et al.} in the context of the KMP model
\cite{Bertini:2007Stochastic}.  

Furthermore, the strict linearity of the temperature profile, which is
linked to the gradient property, is lost for more general sequences of
shape parameters. Provided linearity is recovered at some intermediate
scale, however, one recovers a simple expression of the heat
conductivity in terms of the harmonic mean of the sequence of shape
parameters, similar to the diffusion coefficient of disordered
lattices \cite{Alexander:1981Excitation}.

\appendix

\section{\label{app:1st} Computation of the degree $1$ contributions}  

To derive equation \eqref{eq:KMPbetanfromdt}, replace the argument of
the Laguerre polynomials by a differential operator:
\begin{equation}
  \label{eq:app1st}
  L_{p}(\beta_{n}(\xi_{n} + \eta))
  \rme^{-\beta_{n}(\xi_{n} + \eta)}
  = 
  L_{p}(-\partial_a)
  \rme^{-a\beta_{n}(\xi_{n} + \eta)} \Big|_{a=1}\,,
\end{equation}
and likewise for $L_{q}(\beta_{n\pm1}(\xi_{n\pm1} - \eta))$. Now
compute:
\begin{equation}
  \label{eq:app1st_int}
  \int\rmd\xi_{n} \, \rmd \xi_{n\pm1} \,
  \beta_{n} \beta_{n\pm1} 
  \frac{\xi_{n}}{\xi_{n} + \xi_{n\pm1}} 
  \int_{-\xi_{n}}^{\xi_{n\pm1}}\rmd\eta \,
  \rme^{-a \beta_{n}(\xi_{n} + \eta) - b \beta_{n\pm1} (\xi_{n\pm1} -
    \eta)}  
  = \frac{a \beta_{n} + b \beta_{n\pm1}}
  {2 a^2 b^2 \beta_{n} \beta_{n\pm1}} 
  \,.
\end{equation}
Equation \eqref{eq:KMPbetanfromdt} thus becomes
\begin{equation}
  \label{eq:app1stKMPbetanfromdt} 
  \left.
  L_{p}(-\partial_a)
  L_{q}(-\partial_b)
  \frac{a \beta_{n} + b \beta_{n\pm1}}
  {2 a^2 b^2 \beta_{n} \beta_{n\pm1}} 
  \right|_{a=b=1}
  = 
  \sum_{j = 0}^{p}\sum_{k = 0}^{q}
  \left.
    \frac{\binom{p}{j}\binom{q}{k}}{j! \,k!}
    \partial_{a}^{j}\partial_{b}^{k}
    \frac{a \beta_{n} + b \beta_{n\pm1}}
    {2 a^2 b^2 \beta_{n}\beta_{n\pm1}}
  \right|_{a=b=1}\,.
\end{equation}
Evaluating the
derivatives and using the identities
\begin{equation}
  \label{eq:app1stidbinomial}
  \begin{split}
    & \sum_{j=0}^{p}  \binom{p}{j} (-)^{j} = \delta_{p,0}\,,\\
    & \sum_{j=0}^{p}  \binom{p}{j} (-)^{j}j = - \delta_{p,1}\,,
  \end{split}
\end{equation}
we obtain the announced result.

\section{\label{app:high} Computation of higher degree contributions} 

To compute contributions of degree higher than one, we must consider
the transposition of equation \eqref{eq:app1st_int} to factors of
energy variables raised to arbitrary integer powers,
\begin{align}
  \label{eq:apphigh_int}
  \int
  &
    \rmd \xi_{n} \, \rmd\xi_{n\pm1} \, 
    \beta_{n} \beta_{n\pm1} 
    \frac{\xi_{n}^{c} \xi_{n\pm1}^{d}}{\xi_{n} + \xi_{n\pm1}} 
    \int_{-\xi_{n}}^{\xi_{n\pm1}}\rmd\eta \,
    \rme^{-a \beta_{n}(\xi_{n} + \eta) -
    b \beta_{n\pm1} (\xi_{n\pm1} - \eta)}    \,,
    \cr
  & =
    \int\rmd\xi_{n} \, \rmd\xi_{n\pm1} \,
    \beta_{n} \beta_{n\pm1} 
    \frac{\xi_{n}^{c} \xi_{n\pm1}^{d}}
    {\xi_{n} + \xi_{n\pm1}} 
    \frac{\rme^{-b \beta_{n\pm1} (\xi_{n} + \xi_{n\pm1})}
    - \rme^{-a \beta_{n}(\xi_{n} + \xi_{n\pm1})}}
    {a \beta_{n} - b \beta_{n\pm1}}
    \,,
    \cr
  & = 
    \beta_{n} \beta_{n\pm1}
    \frac{c! \, d!}{c + d + 1}
    \frac{(b \beta_{n\pm1})^{-(c+d+1)} - (a \beta_{n})^{-(c+d+1)} }
    {a \beta_{n} - b \beta_{n\pm1}}
    \,,
    \cr
  & =
    \beta_{n} \beta_{n\pm1}
    \frac{c! \, d!}{c + d + 1}
    \sum_{i = 0}^{c+d} (a \beta_{n})^{-(c+d-i+1)} (b \beta_{n\pm1})^{-(i+1)}
    \,.
\end{align}

To determine the action of the product of Laguerre operators on this
expression,  let $x = a\beta_{n}$ and $y = b \beta_{n\pm1}$
and evaluate the results of derivatives at $x = \beta_{n}$ and $y =
\beta_{n\pm1}$. It is sufficient to consider the $x$-part only, for
which we have  
\begin{align}
  \label{eq:Laguerreaction}
  L_{p}(-\beta \partial_x) x^{-s} 
  &
    = 
    \sum_{j=0}^{p} \frac{\binom{p}{j}}{j!} \beta^{j} \partial_x^j
    x^{-s}
    \,,
    \cr
  &
    = \frac{\beta^{-s}}{(s-1)!}
    \sum_{j=0}^{p} \frac{\binom{p}{j}}{j!} 
    (-)^{j} (s + j - 1)!
    \,,
    \cr
  &
    = 
    \beta^{-s}
    \binom{p-s}{-s}
    \,,
\end{align}
where the binomial factor is $(1-s) \dots (p-s)/p!$ and thus vanishes
for $p \geq s$. 

Finally, the action of the product of the two Laguerre operators on
equation \eqref{eq:apphigh_int} becomes 
\begin{multline}
  \label{eq:apphigh_int2}
  \beta_{n} \beta_{n\pm1}
  \frac{c! \, d!}{c + d + 1}
  L_{p}(-\beta_{n} \partial_x)   L_{q}(-\beta_{n\pm1} \partial_y)
  \sum_{i = 0}^{c+d} x^{-(c+d-i+1)} y^{-(i+1)}
  \\
  = 
  \frac{c! \, d!}{c + d + 1}
  \sum_{i = 0}^{c+d} 
  \beta_{n}^{-(c+d - i)}
  \beta_{n\pm1}^{-i}
  \binom{p-(c+d-i+1)}{-(c+d-i+1)}
  \binom{q-(i+1)}{-(i+1)}
  \,,
\end{multline}
which vanishes identically when $p + q > c+d$ and justifies our
assertion that the coefficients in the expansion of the stationary
state \eqref{eq:KMPness} are determined through sets of equations
which are closed degreewise. 

We proceed to evaluate this expression for the first few degrees.

\subsection{Degree-$0$ term}

Letting $c = d = 0$ in equation \eqref{eq:apphigh_int2}, we get back
the conservation of probability,  
\begin{equation}
  \label{eq:pphigh_o0}
  \delta_{p,0} \, \delta_{q,0}
  \,.
\end{equation}

\subsection{Degree-$1$ terms}

For  $c = 1$ and $d = 0$ (or equivalently $c = 0$ and $d = 1$),  we
retrieve the two contributions to the right-hand side of equation
\eqref{eq:KMPbetanfromdt}, which, up to factor $1/2$, are
\begin{equation}
  \label{eq:pphigh_o1}
  (\beta_{n}^{-1} + \beta_{n\pm1}^{-1}) 
  \delta_{p,0} \, \delta_{q,0}
  -  \beta_{n}^{-1} \delta_{p,1} \, \delta_{q,0}
  -  \beta_{n\pm1}^{-1} \delta_{p,0} \, \delta_{q,1}
  \,.
\end{equation}

\subsection{Degree-$2$ terms}

For  $c = 2$ and $d = 0$ (or $c = 0$ and $d = 2$),  we have, apart
from a factor $2/3$, the contributions 
\begin{multline}
  \label{eq:pphigh_o2}
    (\beta_{n}^{-2} + \beta_{n}^{-1} \beta_{n\pm1}^{-1}
    + \beta_{n\pm1}^{-2}) 
    \delta_{p,0} \, \delta_{q,0}
    - (2\beta_{n}^{-2} + \beta_{n}^{-1} \beta_{n\pm1}^{-1} )  
    \delta_{p,1} \, \delta_{q,0}
    \\
    - (\beta_{n}^{-1} \beta_{n\pm1}^{-1} + 2 \beta_{n\pm1}^{-2} )  
    \delta_{p,0} \, \delta_{q,1}
    + \beta_{n}^{-2} \delta_{p,2} \, \delta_{q,0}
    + \beta_{n}^{-1} \beta_{n\pm1}^{-1}  \delta_{p,1} \, \delta_{q,1}
    + \beta_{n\pm1}^{-2} \delta_{p,0} \, \delta_{q,2}
    \,,
\end{multline}
whose sum yields the right-hand side of equation \eqref{eq:KMPbetan2}.

The contributions corresponding to $c = 1$ and $d = 1$, given by
\eqref{eq:pphigh_o2} multiplied by $1/3$, bring about one
of three contributions to equation \eqref{eq:KMPbetannp1}, namely the
$(n, n+1)$ interaction, the two others, $(n-1, n)$ and $(n+1, n+2)$,
involving corrections of degree $1$ only, given by equation
\eqref{eq:pphigh_o1}. 

\subsection{Degree-$3$ terms}

For  $c = 3$ and $d = 0$ (or $c = 0$ and $d = 3$),  we have, up to a
factor $3/2$ the contributions 
\begin{align}
  \label{eq:pphigh_o3}
  &
    (\beta_{n}^{-3} + \beta_{n}^{-2} \beta_{n\pm1}^{-1}
    + \beta_{n}^{-1} \beta_{n\pm1}^{-2} + \beta_{n\pm1}^{-3}) 
    \delta_{p,0} \, \delta_{q,0}
    - \beta_{n}^{-1} 
    (3 \beta_{n}^{-2} 
    + 2\beta_{n}^{-1} \beta_{n\pm1}^{-1} 
    + \beta_{n\pm1}^{-2} 
    )  
    \delta_{p,1} \, \delta_{q,0}
    \cr
  &
    \quad
    - \beta_{n\pm1}^{-1} 
    (3 \beta_{n\pm1}^{-2} 
    + 2\beta_{n}^{-1} \beta_{n\pm1}^{-1} 
    + \beta_{n}^{-2})  
    \delta_{p,0} \, \delta_{q,1}
    + \beta_{n}^{-2} 
    (3 \beta_{n}^{-1}  + \beta_{n\pm1}^{-1}) 
    \delta_{p,2} \, \delta_{q,0}
    \cr
  &
    \quad
    + 2  \beta_{n}^{-1} \beta_{n\pm1}^{-1} 
    (\beta_{n}^{-1} + \beta_{n\pm1}^{-1}) 
    \delta_{p,1} \,\delta_{q,1}
    + \beta_{n\pm1}^{-2} 
    (\beta_{n}^{-1}  + 3\beta_{n\pm1}^{-1}) 
    \delta_{p,0} \, \delta_{q,2}
    \cr
  &
    \quad
    - \beta_{n}^{-3}
    \delta_{p,3} \, \delta_{q,0}
    - \beta_{n}^{-2} \beta_{n\pm1}^{-1}
    \delta_{p,2} \, \delta_{q,1}
    - \beta_{n}^{-1} \beta_{n\pm1}^{-2}
    \delta_{p,1} \, \delta_{q,2}
    - \beta_{n\pm1}^{-3}
    \delta_{p,0} \, \delta_{q,3}
    \,.
\end{align}
The same contributions hold for  $c = 2$ and $d = 1$ or $c = 1$ and $d
= 2$, up to a factor one half. Altogether these contributions provide
the means to extend our analysis of the stationary state
\eqref{eq:KMPness} to degree $3$.

\section{\label{app:alpha}Extension to parameter-dependent models}

For the model \eqref{eq:KMPalphakernel}, one finds that the right-hand
side of equation \eqref{eq:apphigh_int} transposes to 
\begin{multline}
  \label{eq:appalpha_int}
  \nu \frac{\Gamma(\alpha_{n} + \alpha_{n\pm1})}
  {\Gamma(\alpha_{n})^{2}
  \Gamma(\alpha_{n\pm1})^{2}}
  \frac{\Gamma(\alpha_{n}+c) \Gamma(\alpha_{n\pm1}+d)}
  {\Gamma(\alpha_{n}+\alpha_{n\pm1}+c+d)}
  a^{-\alpha_{n}}\,b^{-\alpha_{n\pm1}}
  \cr
  \times
  \sum_{i=0}^{c+d}
  \binom{c+d}{i}
  \Gamma(\alpha_{n}+c+d-k)
  \Gamma(\alpha_{n\pm1}+k)
  (\beta_{n} a)^{-(c+d-k)}
  (\beta_{n\pm1} b)^{-k}.
\end{multline}
Acting on this expression with the generalised Laguerre
operators $\Ja[0]_{p}(-\partial_{a})$ and $\Ja[0]_{q}(-\partial_{b})$
expanded in closed form, one obtains
\begin{multline}
  \label{eq:appalpha_res}
  \frac{\Gamma(\alpha_{n} + \alpha_{n\pm1})}
  {\Gamma(\alpha_{n}) \Gamma(\alpha_{n\pm1})}
  \frac{\Gamma(\alpha_{n}+c) \Gamma(\alpha_{n\pm1}+d)}
  {\Gamma(\alpha_{n}+\alpha_{n\pm1}+c+d)}
  \sqrt{\frac{\Gamma(\alpha_{n} + p) \Gamma(\alpha_{n\pm1} + q)}
    {p!\,q!\,\Gamma(\alpha_{n}) \Gamma(\alpha_{n\pm1})}}
  \sum_{i=0}^{c+d}
  \beta_{n}^{-(c+d-i)} \beta_{n\pm1}^{-i}
  \binom{c+d}{i}
  \cr
  \times
  \sum_{k = 0}^{p}
  \binom{p}{k}
  (-)^{k}
  \frac{\Gamma(\alpha_{n} + c + d - i + k)}
  {\Gamma(\alpha_{n} + k)}
  \sum_{l = 0}^{q}
  \binom{q}{l}
  (-)^{l}
  \frac{\Gamma(\alpha_{n\pm1} + i + l)}
  {\Gamma(\alpha_{n\pm1} + l)}
  \,.
\end{multline}
For $i$ and $j$ integers, observe that $\Gamma(\alpha + i +
j)/\Gamma(\alpha + j) = (\alpha + j)_{i}$, the Pochhammer symbol, is a
polynomial in $\alpha + j$ of degree $i$. It follows from an algebraic
identity \cite{Ruiz:1996Algebraic} that the corresponding binomial 
series in \eqref{eq:appalpha_res} vanish whenever the indices $k < c +
d -i$ or $l < i$. As a consequence, the right-hand side of
\eqref{eq:appalpha_res} is zero when $p + q>c + d$. Moreover, if $p
\leq i$, the identities \eqref{eq:app1stidbinomial} generalise to
\begin{equation}
  \label{eq:app2ndidbinomial}
  \sum_{j=0}^{p}  \binom{p}{j} (-)^{j} j^{i}  = (-)^{p} p! \, \delta_{p,i}\,.
\end{equation}
Thus, let 
\begin{equation}
  \label{eq:pochhammerexpand}
  (\alpha + k)_{i} = 
  \frac{\Gamma(\alpha + i + k)}
  {\Gamma(\alpha + k)}
  = 
  \sum_{n = 0}^{\infty} C_{n}(\alpha,i) k^{n}
\end{equation}
with the coefficients $C_{n}(\alpha,i) = 0$ if $n > i$. After
substituting these expressions in \eqref{eq:appalpha_res}, we have
\begin{multline}
  \label{eq:appalpha_finalres}
  \nu \frac{\Gamma(\alpha_{n} + \alpha_{n\pm1})}
  {\Gamma(\alpha_{n}) \Gamma(\alpha_{n\pm1})}
  \frac{\Gamma(\alpha_{n}+c) \Gamma(\alpha_{n\pm1}+d)}
  {\Gamma(\alpha_{n}+\alpha_{n\pm1}+c+d)}
  \sqrt{\frac{\Gamma(\alpha_{n} + p) \Gamma(\alpha_{n\pm1} + q)}
    {p!\,q!\,\Gamma(\alpha_{n}) \Gamma(\alpha_{n\pm1})}}
  (-)^{p+q} p!\,q!
  \cr
  \times
  \sum_{i=0}^{c+d}
  \beta_{n}^{-(c+d-i)} \beta_{n\pm1}^{-i}
  \binom{c+d}{i}
  C_{p}(\alpha_{n},\,c+d-i)
  C_{q}(\alpha_{n\pm1},\,i)
  \,.
\end{multline}
Below, we proceed to evaluate this expression for terms of degrees $1$
and $2$.

\subsection{Degree-$1$ terms}

For  $c = 1$ and $d = 0$, up to factor $\nu \alpha_{n}/(\alpha_{n} +
\alpha_{n\pm1})$, we have the three contributions 
\begin{equation}
  \label{eq:appalpha_o1}
  (\alpha_{n} \beta_{n}^{-1} + \alpha_{n\pm1} \beta_{n\pm1}^{-1}) 
  \delta_{p,0} \, \delta_{q,0}
  -  \sqrt{\alpha_{n}} \beta_{n}^{-1} \delta_{p,1} \, \delta_{q,0}
  -  \sqrt{\alpha_{n\pm1}} \beta_{n\pm1}^{-1} \delta_{p,0} \,
  \delta_{q,1}
  \,.
\end{equation}
The addition of the $p=0$ and $q=0$ contributions for the pairs
$\{n-1,n\}$ and $\{n,n+1\}$ minus twice $\alpha_{n} \beta_{n}^{-1}$,
which arises from the lost term in \eqref{eq:KMPmeq}, yield equation 
\eqref{eq:KMPalphabetan}. 

\subsection{Degree-$2$ terms}

Considering $c = 2$ and $d = 0$, up to a common factor 
\begin{equation}
  \label{eq:appalpha_o2fc2d0}
  \nu \frac{ \alpha_{n}(1 + \alpha_{n})}
  {(\alpha_{n} + \alpha_{n\pm1}) (1+ \alpha_{n} + \alpha_{n\pm1})}\,, 
\end{equation}
we have the six contributions 
\begin{multline}
  \label{eq:appalpha_o2}
    \Big\{
    [\alpha_{n}(1 + \alpha_{n}) ]\beta_{n}^{-2}
    + 2\alpha_{n} \alpha_{n\pm1} \beta_{n} \beta_{n\pm1}
    +\alpha_{n\pm1}(1 + \alpha_{n\pm1}) ]\beta_{n\pm1}^{-2}
    \Big\}     \delta_{p,0} \, \delta_{q,0}
    \cr
    -2 \sqrt{\alpha_{n}} \beta_{n}^{-1} 
    [(1 + \alpha_{n}) \beta_{n}^{-1} 
    +\alpha_{n\pm1} \beta_{n\pm1}^{-1} ]  
    \delta_{p,1} \, \delta_{q,0}
    \cr
    -2\sqrt{\alpha_{n\pm1}} \beta_{n\pm1}^{-1} 
    [(1 + \alpha_{n\pm1}) \beta_{n\pm1}^{-1}  
    + \alpha_{n}\beta_{n}^{-1} ]  
    \delta_{p,0} \, \delta_{q,1} 
    + \sqrt{2\alpha_{n}(1 + \alpha_{n})} \beta_{n}^{-2} 
    \delta_{p,2} \, \delta_{q,0}
    \cr
    + 
    2 \sqrt{\alpha_{n} \alpha_{n\pm1}} 
    \beta_{n}^{-1} \beta_{n\pm1}^{-1}  
    \delta_{p,1} \, \delta_{q,1}
    + \sqrt{2\alpha_{n\pm1}(1 + \alpha_{n\pm1})} 
    \beta_{n\pm1}^{-2} \delta_{p,0} \, \delta_{q,2}
    \,.
\end{multline}
The same expression \eqref{eq:appalpha_o2} is obtained for $c=0$ and
$d=2$, except for the factor \eqref{eq:appalpha_o2fc2d0}, which is
replaced by
\begin{equation}
  \label{eq:appalpha_o2fc0d2}
  \nu \frac{ \alpha_{n\pm1}(1 + \alpha_{n\pm1})}
  {(\alpha_{n} + \alpha_{n\pm1}) (1+ \alpha_{n} + \alpha_{n\pm1})}\,, 
\end{equation}
as well as for $c=1$ and $d=1$, with the factor
\eqref{eq:appalpha_o2fc2d0} replaced by
\begin{equation}
  \label{eq:appalpha_o2fc1d1}
  \nu \frac{ \alpha_{n} \alpha_{n\pm1}}
  {(\alpha_{n} + \alpha_{n\pm1}) (1+ \alpha_{n} + \alpha_{n\pm1})}\,.
\end{equation}
Combinations involving these terms lead to the expressions
\eqref{eq:KMPalphauniformbetan2}-\eqref{eq:KMPalphauniformbetannp1}
and \eqref{eq:KMPalpha01betan2}-\eqref{eq:KMPalpha01betannp1}. 

\ack{The author wishes to acknowledge the hospitality of the Erwin
  Schr\"odinger Institute, Vienna, on the occasion of the conference
  Hyperbolic Dynamics and Statistical Physics held in May 2016, where
  part of this work was presented. He receives financial support from
  the (Belgian) FRS-FNRS. 
}

\section*{References}

\bibliography{kmprevisited.bbl}

\end{document}